\documentclass[11pt,a4paper]{article} 
\pdfoutput=1
\usepackage{jcappub}
\usepackage{multirow}

\newcommand{\nc}{\newcommand}
 
\nc{\lcdm}{$\Lambda$CDM }
\nc{\lya}{L$\alpha$F }
\nc{\1}{\aa}
\nc{\2}{\"{a}}
\nc{\3}{\"{o}}
\nc{\4}{\AA}
\nc{\5}{\"{A}}
\nc{\6}{\"{O}}
\nc{\be}[1]{\begin{equation}\mbox{$\label{#1}$}}
\nc{\bea}[1]{\begin{eqnarray} \mbox{$\label{#1}$}}
\nc{\eea}{\end{eqnarray}}
\nc{\ee}{\end{equation}}

\title{\bf Can a void mimic the $\Lambda$ in $\Lambda$CDM?}
\author[a,b]{Peter Sundell,}
\author[c,d]{Edvard M\3rtsell}
\author[a]{and Iiro Vilja}
\affiliation[a]{Turku Center for Quantum Physics, Department of Physics and Astronomy, \\
University of Turku,   FIN-20014 Turku, Finland}
\affiliation[b]{Nordita, KTH Royal Institute of Technology and Stockholm University, \\ 
Roslagstullsbacken 23, SE-10691 Stockholm, Sweden}
\affiliation[c]{The Oskar Klein Centre for Cosmoparticle Physics, Stockholm University\\
AlbaNova University Center, SE 106 91 Stockholm, Sweden}
\affiliation[d]{Department of Physics, Stockholm University\\
AlbaNova University Center, SE 106 91 Stockholm, Sweden}
\date{\today}

\emailAdd{pgsund@utu.fi}
\emailAdd{edvard@fysik.su.se}
\emailAdd{iiro.vilja@utu.fi}

\abstract{
We investigate Lema\^{i}tre-Tolman-Bondi (LTB) models, whose early time evolution and bang time are homogeneous and the distance - redshift relation and local Hubble parameter are inherited from the \lcdm model.  We show that the obtained LTB models and the  \lcdm model predict different relative local expansion rates and that the Hubble functions of the models  diverge increasingly with redshift. The LTB models show tension between low redshift baryon acoustic oscillation and supernova observations  and including Lyman-$\alpha$ forest or  cosmic microwave background observations only accentuates the better fit of the \lcdm model compared to the LTB model. The result indicates that additional degrees of freedom are needed to explain the observations, for example by renouncing spherical symmetry,  homogeneous bang time,  negligible effects of pressure, or the early time homogeneity assumption.}


\begin{document}
\maketitle

\section{Introduction}

Supernova Ia (SNIa) observations  at the end of the 1990's contain the peculiar feature that the universe appears to be accelerating \cite{Riess1998,Perlmutter1999}. In spite of years passing by, the phenomenon remains without complete understanding; the issue is generally recognised as the dark energy problem. Attempts to solve this problem include modifying gravity and relaxing the cosmological principle, but so far has no single theory explained the accelerating expansion together with the other cosmological phenomena better than the \lcdm model. 

In the framework of Einstein's general relativity, a natural way to create apparent acceleration is to use large-scale inhomogeneous matter distributions \cite{Celerier}, which means giving up the cosmological principle. As the principle is not observationally confirmed \cite{VMC2012,RedlichEtAl2014}, renouncing it is merely a philosophical question and as such a widely considered option. However, to describe an inhomogeneous matter distribution in general relativity is mathematically a very challenging task, e.g., the linear perturbation theory of  the  Lema\^{i}tre-Tolman-Bondi (LTB) models \cite{Lemaitre,Tolman,Bondi} is not completely understood \cite{ClarksonCliftonFebruary2009}, even though the LTB models are considered as simple inhomogeneous models. This challenge is to be taken step by step, revealing the possibilities brought by inhomogeneously distributed matter starting from the simplest models.

A vastly investigated LTB model is where the observer is located inside a void.  It is often assumed that the void is embedded in a homogeneous and isotropic background, and that the void was shallow enough to be indistinguishable from its background at early times. This simplification allows one to apply homogeneous and isotropic perturbation theory.  The observed isotropy of the cosmic microwave background (CMB) enforces the observer to be located very close to the center of the void \cite{AlnesAmarzguioui2006,BlomqvistMortsell2010}. Attempts have been made to find models where this property is removed, but it appears that the LTB models do not have enough degrees of freedom to allow it \cite{SundellVilja2014}. 

LTB models, where the observer is located at the symmetry center, are widely studied. The authors of Ref.~\cite{Garcia-BellidoHaugbolle2008a} found SNIa, first CMB peak and baryon acoustic oscillation (BAO) observations to fit the LTB model  they used. In Ref.~\cite{BNV2010}, their model was able to provide good fits to the same kind of data, but encountered difficulties with the local value of the Hubble parameter ($H_0$). In the models used in Ref.~\cite{ZG-BR-L2012}, the authors found strong tension between SNIa and BAO observations, which became even worse when  CMB observations was included. Additional problems were found for the void that was embedded in a spatially flat Friedmann-Robertson-Walker (FRW) background when $H_0$ was included, but they were removed by allowing the background to be hyperbolically curved. The authors in Ref.~\cite{MossZibinScott2011} found their models  to fit to the full CMB spectrum and SNIa data, but the models suffered from a low $H_0$. Similar tensions between SNIa, $H_0$ and CMB observations were encountered in Ref~\cite{RedlichEtAl2014}. The LTB models can explain the kinematic Sunyaev-Zeldovich (kSZ) effect \cite{Garcia-BellidoHaugbolle2008b} only if  SNIa, CMB and $H_0$ fits are not considered \cite{ZibinMoss2011,ZhangStebbins2011,BCF2012}. The results from studies subjecting LTB models to observational data thus do not paint a clear picture of the pros and cons of the models. However, since different authors use different models, data sets and methods, the results are not necessarily contradictory.

Our aim is to bring a different understanding to these results.  We fix the luminosity distance - redshift relation to be equivalent with that of the \lcdm model. Then, we investigate which quantities differ most from their \lcdm counterparts, and study observables which are dependent on the most diverging quantities.  This type of LTB model has been studied before \cite{IguchiNakamuraNakao2002}, but the issues at the apparent horizon (AH) have only recently been resolved \cite{Krasinski2014} making it more appealing to study the model further.\footnote{The AH is the location where the derivative of the angular diameter distance with respect to the coordinate distance vanishes.} A disadvantage of this LTB model, where the observer is at the symmetry center and the bang time is homogeneous,  is that it can mimic the distance - redshift relation only up to redshift  $z\approx 6.9$ \cite{Krasinski2014} after which further properties must be described by a model which no longer has the same distance - redshift relation as the \lcdm model.

LTB models are often analysed in terms of their specific relation to observations. This type of studies give us apprehension of the properties of the LTB models but not a full understanding about its failures and advantages. However, the fact that we have some sort of comprehension of the models capabilities, gives us a direction to go when generalising to more complex models; if we know where the LTB models appear to fail, it is wise to look carefully into these issues.

In Section \ref{LTB}, we introduce the general properties of the LTB models relevant to this work and in Section \ref{system}, we derive the  differential equations of the LTB model with homogeneous bang time, the  \lcdm distance - redshift relation and $H_0$  in notation similar to that usually used in the \lcdm model. In Section \ref{results}, we give an overview of how we solve the relevant differential equations and present the solutions; technical details of the procedure are given in Appendices \ref{AA}-\ref{AAH}. In the  following Section, \ref{comparison}, we compare quantities predicted by the obtained model with those predicted by the \lcdm model. On the strength of these results, we continue by comparing these models with BAO (including Lyman-$\alpha$ forest) and CMB observations. We end the article with conclusions and a discussion in Section \ref{conclusions}.

\section{The  Lema\^{i}tre-Tolman-Bondi models} \label{LTB}

The LTB models describe a spherically symmetric dust distribution. The metric is
\be{m}
ds^2=-dt^2+\frac{R_{,r}^2(r,t)}{1+2 e(r) r^2} dr^2+R^2(r,t)(d\theta^2+\sin^2 \theta d\phi^2)
\ee
and the Einstein equations yield the evolution equation
\be{Rt^2}
R_{,t}^2(r,t)=\frac{2M(r)}{R(r,t)}+2e(r)r^2,
\ee
where the functions $M(r)$ and $e(r)$ are time independent integration constants. The equation for the physical matter distribution is
\be{rho}
\rho(r,t)=\frac{2 M'(r)}{\kappa R^2(r,t) R_{,r}(r,t)},
\ee
where $\kappa=8 \pi G/c^2$. We denote  partial derivatives with subscript $r$ or $t$ after commas. The solution of the former equation can be given in parametric form,
\be{R}
R(r,t)=\frac{M(r)}{2e(r) r^2}\left\{\cosh[\eta(r,t)]-1\right\},
\ee
\be{eta}
\sinh[\eta(r,t)]-\eta(r,t)= \frac{[2e(r)r^2]^{3/2}}{M(r)}[t-t_b(r)],
\ee
where $t_b(r)$ is the bang time function.

The redshift, $z$, and the coordinate distance, $r$, are related by
\be{rs}
\frac{dz}{dr}=(1+z)\frac{R_{,tr}(r,t)}{\sqrt{1+2e(r)r^2}},
\ee 
where $t$ should satisfy the null geodesic equation for incoming light rays
\be{ng}
\frac{dt}{dr}=-\frac{R_{,r}(r,t)}{\sqrt{1+2e(r)r^2}}.
\ee
Using Eq.\ \eqref{R}, the functions $R_{,r}(r,t)$ and $R_{,tr}(r,t)$ can be given as functions of $M(r)$, $e(r)$, $r$ and $\eta(r,t)$. The dependence on the conformal time, $\eta(r,t)$, can be eliminated using Eq. \eqref{eta}, yielding
\bea{Rr}
R_{,r}(r,t)&=&\left[-\frac{\left(2 r^2 e'(r)+4 r e(r)\right)}{2 r^2 e(r)}+\frac{M'(r)
   }{M(r)}\right] R(r,t) \nonumber \\
&&+ \left[\frac{3 \left(2 r^2 e'(r)+4 r e(r)\right)
   (t-t_b(r))}{4 r^2 e(r)}-\frac{M'(r) (t-t_b(r))}{M(r)}-t_b'(r)\right]R_{,t}(r,t), \\
R_{,tr}(r,t)&=&R_{,t}(r,t) \left[\frac{e'(r)}{2 e(r)}+\frac{1}{r}\right] \nonumber \\ \label{Rtr}
&&-\frac{M(r)}{R(r)^2} \left[\frac{3 \left(2 r^2 e'(r)+4 r e(r)\right) (t(r)-t_b(r))}{4 r^2 e(r)}-\frac{M'(r)
   (t(r)-t_b(r))}{M(r)}-t_b'(r)\right]. \quad
\eea

The apparent horizon (AH) is defined as the locus where $dR[t(r),r]/dr=0$, where $t(r)$ is given by the null geodesic equation \eqref{ng}  \cite{BKHC}. This definition reduces to the form
\be{ah}
R[r,t(r)]=2 M(r), 
\ee
which can be seen by using the null geodesic equation \eqref{ng}.

To ensure that the positive energy condition, $\rho(r,t)>0$, is fulfilled, we demand $M'(r)\geq 0$ and $R_{,r}(t,r)\geq 0$ throughout this paper (see  Eq.\ \eqref{rho}). Furthermore, to avoid shell crossing (SC) singularities, defined as surfaces where $R_{,r}(r,t)=0$ and $M'(r)\neq 0$, inequalities $t'_b(r)\leq 0$ and
\be{sc}
 \frac{d}{dr}\left[e(r) r^2 \right] \geq 0
\ee
should always be satisfied \cite{HellabyLake1985}.

\section{The Lema\^{i}tre-Tolman-Bondi models mimicking the \lcdm model}\label{system}

We consider LTB models where the observer is located at the center, the bang time function is constant, and the Hubble constant and luminosity distance - redshift relation are equal to those in the \lcdm model.   Thus, a LTB model automatically gives fits identical to the corresponding \lcdm model to $H_0$,  SNIa  and CMB dipole observations.

Requiring that the LTB and the \lcdm luminosity distances are equal is equivalent to requiring the angular diameter distances to be equal \cite{IguchiNakamuraNakao2002}. The \lcdm angular diameter distance is given by
\be{add}
D_A=\frac{1}{H_0^F(1+z)\sqrt{\Omega_k^F}}\sinh\left\{\sqrt{\Omega_k^F}\int_0^z\frac{d\tilde{z}}{\sqrt{\Omega_m^F(1+\tilde{z})^3+\Omega_k^F(1+\tilde{z})^2+\Omega_w^F(1+\tilde{z})^w+\Omega_{\Lambda}^F}}\right\},
\ee
where $H_0^F$ is the ``Friedmannian'' Hubble parameter and $\Omega_m^F$, $\Omega_k^F$, $\Omega_{\Lambda}^F$ the corresponding matter, curvature, and cosmological constant densities, respectively. In addition to a cosmological constant, we have included a constant equation of state ($\rho= w p$) term $\Omega_w^F$.  The standard relation $\Omega_m^F+\Omega_k^F+\Omega_w^F+\Omega_{\Lambda}^F=1$ holds.

The angular diameter distance for the  LTB models is $R[r(z),t(z)]$, thus imposing
\be{add2}
R[r(z),t(z)]=D_A,
\ee
the LTB models have the same angular diameter distance as the corresponding  \lcdm models.
To ensure that the models behave as a homogeneous and isotropic models at early times, we set the bang time to a constant, which without loss of generality can be set to zero, i.e.
\be{tb}
t_b(r)=0.
\ee 
This choice immediately satisfies the condition $t'_b(r)\leq 0$ required to avoid SC singularities, thus the only condition left to be satisfied to prevent SC singularities is Eq.\ \eqref{sc}.

In order to write Eq.\ \eqref{Rt^2} in a more familiar form, we define
\bea{defs}
  M(r)&=&\frac{1}{2}\Omega_M(r) H_0^2(r) R_0^3(r), \nonumber \\
   e(r)r^2&=&\frac{1}{2}\Omega_K(r) H_0^2(r) R_0^2(r), \nonumber \\
H_R(r,t)&=&\frac{R_{,tr}(r,t)}{R_{,r}(r,t)},  \\
H_T(r,t)&=&\frac{R_{,t}(r,t)}{R(r,t)}, \nonumber \\
H_0(r)&=&H_T(r,t_0), \nonumber
\eea
where $R_0(r)=R(r,t_0)$ and $t_0$ is the present time. Two Hubble functions are defined  above: $H_T$  is the transverse Hubble function and $H_R$  is the radial Hubble function.  The geometrical mean of these Hubble functions in the three spatial directions is $H_{LTB}(r,t)=\left[H_R(r,t) H_T^2(r,t)\right]^{1/3}$.
These definitions allow us to rewrite Eq.\ \eqref{Rt^2} as
\be{H}
H_T^2(r,t)=H_0^2(r)\left[\Omega_M(r)\left(\frac{R_0(r)}{R(r,t)}\right)^3+\Omega_K(r)\left(\frac{R_0(r)}{R(r,t)}\right)^2\right].
\ee
Because, at $t_0$ for every $r$, $H_T(r,t_0)=H_0(r)$ and $R(r,t_0)=R_0(r)$, the relation
\be{Omegak}
\Omega_K(r)=1-\Omega_M(r)
\ee
must hold for every $r$ at any given time. Using the  gauge 
\be{gauge}
R_0(r)=r, 
\ee
 Eq.\ \eqref{H} becomes
\be{H2}
H_T^2(r,t)=H_0^2(r)\left[\Omega_M(r)\left(\frac{r}{R(r,t)}\right)^3+(1-\Omega_M(r))\left(\frac{r}{R(r,t)}\right)^2\right],
\ee
and at the homogeneous limit, where $R(r,t)=a(t)r$ and $\Omega_M(r)$ and $H_0(r)$ are constants, Eq.\ \eqref{H2} reduces to the Friedmann equation.

Using the origin conditions given in Appendix \ref{AB}, we find $H_{LTB}(0,t_0)=H_T(0,t_0)=H_R(0,t_0)=H_0^F$. Thus, to have appropriate conditions at the origin, the LTB models and the corresponding \lcdm models must share the same local Hubble value, namely $H_0^F$.

We will solve the system along null geodesics, where the derivative of $R[r,t(r)]$ is
\be{dRdr}
\frac{d}{dr}R=R_{,r}+R_{,t}t_{,r}=R_{,r}-R_{,t}\frac{R_{,r}}{\sqrt{1+2e(r)r^2}},
\ee
and the second equality is obtained from Eq.\ \eqref{ng}.

The system could be solved using Eqs.\  \eqref{rs}, \eqref{ng}, \eqref{add2} and \eqref{dRdr}, which can be presented as a set of differential equations for $R$, $t$, $z$ and $\Omega$.
However,  in order to improve the numerical stability of the solutions, we  manipulate the set of differential equations into a more convenient form, see Appendix \ref{AA}.

\section{The solution of the system}\label{results}

\begin{figure}[h]
\centering
\includegraphics[scale=0.9]{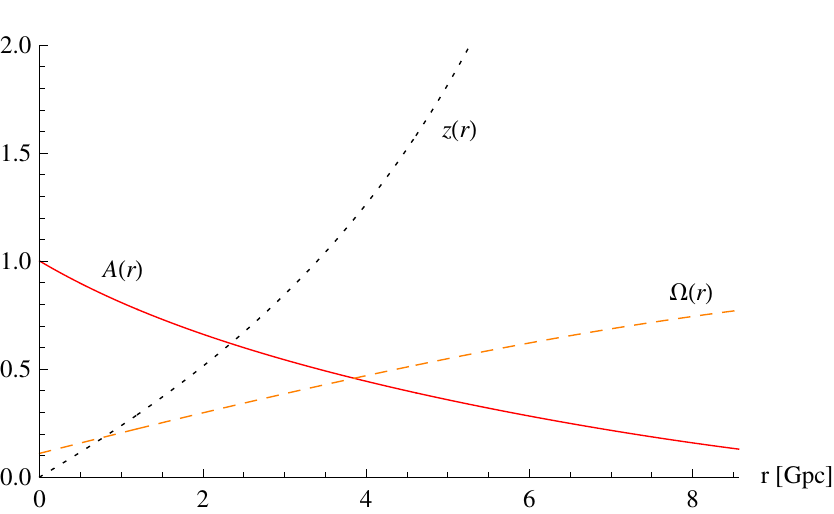} 
\includegraphics[scale=0.9]{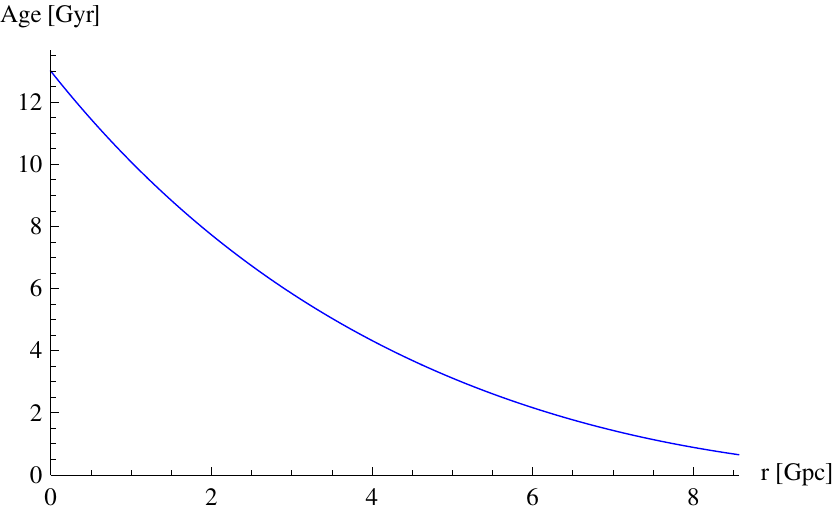} 
\caption{\footnotesize The LTB model with $H_0^F=67.1$ and $\Omega_m^F=0.32$. In the left panel, pieces of the functions,   $\Omega_M$ (dashed orange), $A=R/r$ (solid red) and $z$ (dotted black), are depicted. Time, $t$,  is plotted  in the right panel.  Fits appear perfect, and the redshift takes the value $z\approx 6.9$ at the end point $r=r_{SC}\approx 8.6$. Colors available online. } 
\label{k9}
\end{figure}

In Ref.\ \cite{Krasinski2014}, the system of equations in the special case where $H_0^F$ and $\Omega_m^F$ are fixed is studied. We study the system by allowing  $H_0^F$ and $\Omega_m^F$  to vary. The set of equations solved in \cite{Krasinski2014} differs from ours in two respects; the differential equations are constructed for different quantities and the gauges are chosen differently. We present here only an overview of the issues encountered solving the equations and direct the reader to Appendices \ref{AA}-\ref{AAH} or \cite{Krasinski2014} for more details. We encountered similar difficulties and numerical errors of the same order as those in \cite{Krasinski2014}. 

The most challenging numerical task for numerically solving the system at hand is to find initial conditions so that the condition \eqref{ah} holds at the AH. An algebraic  relation between the initial and the AH conditions can not be found, hence one needs to choose initial conditions and numerically evaluate the system from the origin to the AH and check if the AH condition \eqref{ah}  holds. To preserve numerical accuracy at the origin and the AH, we use linear approximations near these points.

The most accurate solution to the system  is obtained by integrating the system from the vicinity of the origin to an intermediate radius (in  Figure \ref{k9} at $r=1.1$ Gpc) and from the vicinity of the AH (in  Figure \ref{k9} at $r=r_{AH}\approx 4.5$ Gpc) to the same intermediate radius and then gluing these solutions together. Beyond the AH, we continued the integration of the system normally. The quantities $t$, $R/r$, $z$ and $\Omega$ of the glued together solutions of the LTB model with $H_0^F=67.1$ and $\Omega_m^R=0.32$ are shown in Figure \ref{k9}. 

The plots in Figure \ref{k9} ends at $r=r_{SC}\approx 8.6$ Gpc, because the system violates the SC condition \eqref{sc} there. Consequently, the LTB model can not describe a physically acceptable universe at this point when the \lcdm luminosity distance with respect to redshift is imposed. However, the SC takes place at $z\approx 6.9$, where the \lcdm model is dominated by matter. This allows us to continue the system further without adopting the luminosity distance  of the \lcdm model. For our needs, it is sufficient to continue the system beyond the SC singularity by not breaking the SC condition and  by demanding early time homogeneity.  Thus, one should evaluate the physical matter density $\rho(r,t)$ in the early times and make sure it is (approximately) independent of $r$. In this article, however, we use a weaker condition where $\Omega(r)$ is constant from the bang time to the drag epoch. This appears to be adequate for our needs.

\section{Comparison to $\Lambda$CDM model} \label{comparison}

Our model is designed to inherit the luminosity distance from the  \lcdm model   before the SC singularity.  Hence, if some other quantities  differ,  observables dependent on these quantities may differ  too. Therefore,  first we  study which quantities differ considerably between the models and then observables dependent on those quantities. For this, we use the LTB and the \lcdm models with $H_0^F=67.1$ and $\Omega_m^F=0.32$.

The difference between the age functions, $t(z)$,  is interesting, because the age of the universe has been used to constrain the LTB models in the literature. We find that the relative difference is approximately $6\, \%$, $12\,  \%$ and $17\,  \%$, at redshifts $0$, $1$ and $6$, respectively.  This has been studied also in Refs. \cite{MossZibinScott2011} and \cite{dePutterVerdeJimenez2012}.

\begin{figure}[t]
\centering
\includegraphics[scale=0.9]{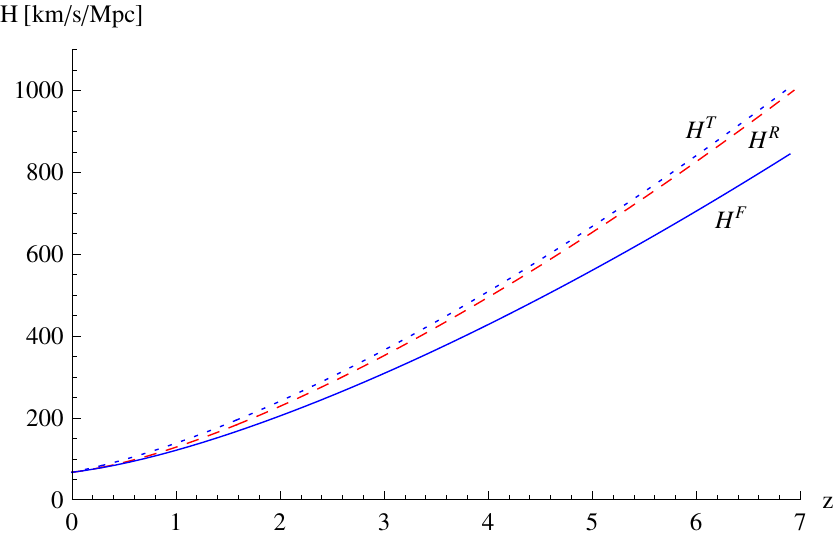} 
\includegraphics[scale=0.9]{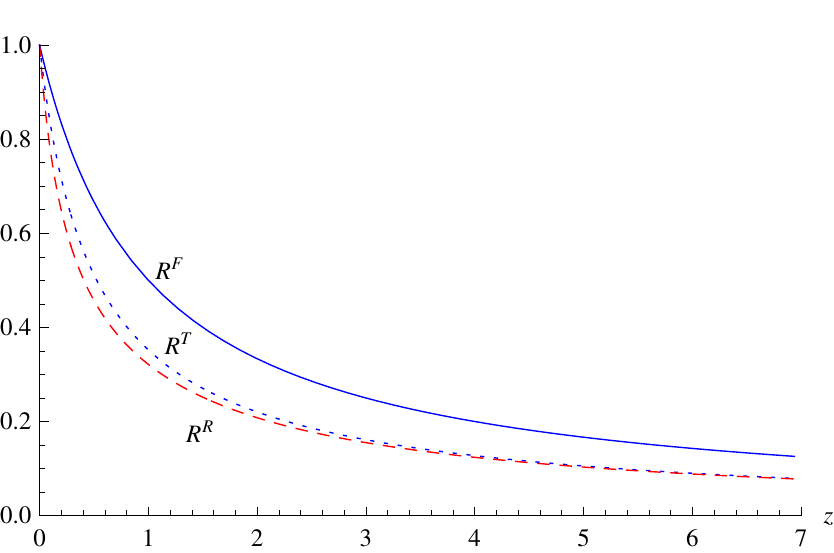} 
\caption{\footnotesize The LTB and the  \lcdm models with $H_0^F=67.1$ and $\Omega_m^F=0.32$. The left panel shows the radial Hubble function, $H^R(z)$ (dashed red),  the transverse  Hubble function, $H^T(z)$ (dotted blue),   and the Hubble function of the \lcdm model, $H^F(z)$ (solid blue). All three Hubble functions coincides at $z=0$. The relative difference between $H^F(z)$ and $H^R(z)$ is approximately  $6\,  \%$ and $16\,  \%$, and the relative difference between $H^F(z)$ and $H^T(z)$ is approximately $13\,  \%$ and $18\,  \%$, at redshifts $1$ and $6$, respectively. The right panel shows RLER's, where $\mathcal{R}^R$, $\mathcal{R}^T$ and $\mathcal{R}^F$ correspond to dashed red, dotted  blue and solid blue curves, respectively.  The relative difference between $\mathcal{R}^F$ and $\mathcal{R}^R$ is approximatelly  $43\,  \%$ and $47\,  \%$, and the relative difference between $\mathcal{R}^F$ and $\mathcal{R}^T$ is approximately $35\,  \%$ and $45\,  \%$, at redshifts $1$ and $6$, respectively. Colors available online.}
\label{k11}
\end{figure}

Instead of investigating quantities $R_{,r}(z)$, $R_{,t}(z)$ and $R_{,tr}(z)$ separately, we investigate the transverse Hubble function $H^T(z)$ and the radial Hubble function $H^R(z)$. The differences between $H^T(z)$, $H^R(z)$ and the \lcdm Hubble function $H^F(z)$ are of similar size as the age function; at the origin, $H^F_0=H^F=H^R=H^T$ and at redshift 6, the   difference between $H^F(z)$ and $H^R(z)$ is approximately $16\,  \%$, and  between $H^F(z)$ and $H^T(z)$ approximately $18\,  \%$, see Figure \ref{k11}.

In the inhomogeneous models,  the expansion depends on position, which is not the case in homogeneous and isotropic models. Hence, the differences between the expansion rates of the \lcdm and the LTB models are expected. From the metric \eqref{m}, one can see that the expansion of the universe  in the radial and transverse directions is encoded  in the functions $R_{,r}(r,t)$ and $R(r,t)$, respectively. An illustrative way to compare expansion rates is to evaluate  how much  $R_{,r}(r,t)$ and $R(r,t)$ have changed  from some early time $t_e$ to the present. For this purpose, one should evaluate the ratios  $R_{,r}[r(z),t(z)]/R_{,r}[r(z),t_e]$, $R[r(z),t(z)]/R[r(z),t_e]$ and $a[t(z)]/a(t_e)$ at different radial coordinates, where $a(t)$ is the scale factor of the \lcdm model. However, these ratios contain the inconvenient  dependence on $t_e$, for which reason we compare the ratios
\bea{ratios5}
\mathcal{R}^R&=&\frac{R_{,r}[r(z_1),t(z_1)]}{R_{,r}[r(z_1),t_e]} \frac{ R_{,r}[r(z_2),t_e]}{R_{,r}[r(z_2),t(z_2)]}, \nonumber \\ \mathcal{R}^T&=&\frac{R[r(z_1),t(z_1)]}{R[r(z_1),t_e]} \frac{ R[r(z_2),t_e]}{R[r(z_2),t(z_2)]}, \\ \mathcal{R}^F&=&\frac{a[t(z_1)]}{a(t_e)} \frac{ a(t_e)}{a[t(z_2)]}. \nonumber 
\eea
 We fix the second factor of the ratios by evaluating it at the origin, hence 
\bea{expR}
\mathcal{R}^R&=&\frac{R_{,r}[r(z),t(z)]}{R_{,r}[r(z),t_e]} A(0,t_e), \\
\mathcal{R}^T&=& \frac{A[r(z),t(z)]}{A[r(z),t_e]} A(0,t_e), \label{expT} \\
\mathcal{R}^F&=&\frac{a[t(z)]}{a(t_e)} \frac{a(t_e)}{a(0)}=\frac{1}{1+z}\label{expF},
\eea
as $R(r,t)=A(r,t) r$, $  R_{,r}(0,t_e)=A(0,t_e)$ and  $R_{,r}(0,t_0)= A(0,t_0)=1$.
The relative local expansion rates (RLER), $\mathcal{R}^R$, $\mathcal{R}^T$, and $\mathcal{R}^F$,   represent the  amount of local expansion from $t_e$ to our null cone compared  to the expansion at the origin. Going through the coordinate values from $r(z=0)$ to $r(z=6.9)$, we obtain  information about how the universe has grown at different coordinate distances. RLER's $\mathcal{R}^R$, $\mathcal{R}^T$ and $\mathcal{R}^F$ of the \lcdm and the LTB models are drawn  in Figure \ref{k11}. Because the amount of local expansion is scaled with the amount of expansion in the origin, $\mathcal{R}^R$, $\mathcal{R}^T$ and $\mathcal{R}^F$ all  take the unit value at $r=0$. Moreover, $\mathcal{R}^R$, $\mathcal{R}^T$ and $\mathcal{R}^F$ all  decrease with radius, showing that the expansion of the LTB model has been fastest at the origin. There are two reasons for this; distant objects have had less time to expand and the surrounding matter distribution is denser at large $r$.

While the $t_e$ dependence of $\mathcal{R}^F$ vanish completely,  $\mathcal{R}^R$, $\mathcal{R}^T$ still explicitly include the parameter $t_e$.  For practical calculations we set $t_e=3 \times 10^{-7}$ (corresponding to about 300 000 years). For this value, the difference between $\rho[r(z=0),t_e]$ and $\rho[r(z=6),t_e]$ is $\sim 10^{-3}\, \%$.

\subsection{Baryon acoustic oscillations}\label{BAO}

The results above indicate that the predictions for observables dependent on RLER's and Hubble functions will differ considerably between the models. As we shall see, baryon acoustic oscillations are such observables. Here we compare the models using not only the  low redshift galaxy BAO measurements, but also  baryon acoustic features obtained from the Lyman-$\alpha$ forest (L$\alpha$F).

\subsubsection{Maximum likelihood analysis}

The characteristic size of baryon acoustic oscillations is given by
\be{dz}
d_{z_{BAO}}=\left( \Delta \theta^2_{BAO} \frac{\Delta z_{BAO}}{z_{BAO}}\right)^{1/3},
\ee
which with the line element \eqref{m} and the redshift equation \eqref{rs}  gives
\be{dzLT}
d_{z_{BAO}}^{LTB}=\left(\frac{( l^T_{phys,BAO})^2}{R_{BAO}^2} \frac{l_{phys,BAO}^R(1+z_{BAO})R_{,tr}^{BAO}}{z_{BAO}R_{,r}^{BAO}}\right)^{1/3}
\ee
 at the lowest order of approximation.  The subscript or superscript BAO indicates that the quantity is evaluated at $z_{BAO}$.

To obtain BAO predictions for the LTB model, we need to evaluate the physical lengths at early times.
As shown e.g. in Ref. \cite{ZG-BR-L2012}, in the LTB model, the physical scales in transverse, $l_{phys}^T$ and radial, $l_{phys}^R$, directions change as
\bea{l}
l_{phys}^T(r,t)&=&\frac{R(r,t)}{R(r,t_{e})} l_{phys}^T(r,t_{e}), \\
l_{phys}^R(r,t)&=&\frac{R_{,r}(r,t)}{R_{,r}(r,t_{e})} l_{phys}^R(r,t_{e}).
\eea
 We require the LTB model to be almost homogeneous at early times. For this reason, we assume that the mean primordial perturbation have grown equally in all directions, i.e., $ l_{phys}^R(r,t_{e})\approx l_{phys}^T(r,t_{e})$. Hence we can collectively denote physical lengths in each direction by  $l_{phys}(r,t_e)$. Furthermore, it is also reasonable to expect the mean perturbation to have the same size everywhere, i.e., $l_{phys}(r_{BAO},t_{e})\approx l_{phys}(r,t_{e})$ for any $r$.\footnote{The early times homogeneity assumption reduces here the degrees of freedom. For the inhomogeneous early time cosmology, however, it would be difficult to determine the functions $ l_{phys}^R(r,t)$ and  $l_{phys}^T(r,t)$. On the other hand, we could use the BAO obsevations to determine these functions.}  Consequently, Eq. (\ref{dzLT}) can be written as
\bea{dzLT2}
d_{z_{BAO}}^{LTB}&=&\left(\frac{1+z_{BAO}}{(D_A^{BAO})^2} \frac{H^R_{BAO}}{z_{BAO}}\right)^{1/3}\left(\frac{R_{,r}^{BAO}}{R_{,r}(r_{BAO},t_e)}\right)^{1/3}\left(\frac{R_{BAO}}{R(r_{BAO},t_e)}\right)^{2/3} l_{phys}(t_e),
\eea
where  Eq.\ \eqref{add2} and the definition for the radial Hubble function in Eq.\  \eqref{defs} are employed. The $\Lambda$CDM counterpart  of the above equation is
\bea{dzf2}
d_{z_{BAO}}^{F}&=&\left(\frac{1+z_{BAO}}{(D_A^{BAO})^2} \frac{H^F_{BAO}}{z_{BAO}}\right)^{1/3} \frac{1+z_{e}}{1+z_{BAO}} l_{phys}(t_e).
\eea

 \begin{table} \centering
 \begin{tabular}{ | c | c | c | c | c | c |}
    \hline
        redshift &  $d_z^{obs}$ & $\Delta z^{obs}$&  $(1+z)\Delta \theta^{obs}$ &  Ref. & abbreviation\\ \hline
    \parbox[t]{0.9cm}{\centering{0.106} 0.35 0.44 0.57 0.60 0.73}   &   \parbox[t]{2.5cm}{ \centering{$ 0.336 \pm 0.015$} $   0.113 \pm 0.0022  $ $0.0916 \pm 0.0071$ $0.073 \pm 0.0012$ $ 0.0726 \pm 0.0034$ $0.0592 \pm 0.0032$}  &  & & \parbox[t]{0.6cm}{\cite{BAO1} \cite{BAO2} \cite{BAO3} \cite{BAO4} \cite{BAO3} \cite{BAO3}} &  \parbox[t]{1.6 cm}{\vspace{0.7 cm}  low or  low BAO} \\ \hline
     \parbox[t]{0.9cm}{\centering{2.34} 2.34 2.36 2.36}   &    &  \parbox[t]{2.4cm}{ $ 0.109 \pm 0.0033  \vspace{0.5 cm}$  $ 0.111 \pm 0.0037$} &  \parbox[t]{2.7cm}{\vspace{0.3 cm} $ 0.08865 \pm 0.0051 \vspace{0.5 cm}$ \centering{$ 0.0926 \pm 0.0.003$}}  &  \parbox[t]{0.6cm}{\cite{BAO5} \cite{BAO5} \cite{BAO6} \cite{BAO6}} &  \parbox[t]{1.5cm}{  L$\alpha$FacR  L$\alpha$FacT \\ L$\alpha$FccR \\ L$\alpha$FccT } \\ \hline 
\end{tabular}
\caption{\footnotesize The BAO observables used in this work. The observational values in Refs. \cite{BAO2,BAO4,BAO5,BAO6} are reported as $ 1/d_{0.35} = 8.88 \pm 0.17$,   $ 1/d_{0.57} =13.67 \pm 0.22$, $1/\Delta z = 9.18 \pm 0.28$,  $[(1+z)\Delta \theta]^{-1}=11.28 \pm 0.65$,   $1/\Delta z =9.00 \pm 0.30$ and  $[(1+z)\Delta \theta]^{-1}= 10.8 \pm 0.4$, respectively, but on the table we present their inverses. In the table, measurements are devided into low \cite{BAO1,BAO2,BAO3,BAO4} and high \cite{BAO5,BAO6} redshift BAO measurements.  } 
  \label{dzobs} \end{table}

 We present the results of the high redshift BAO (or L$\alpha$F)  surveys  \cite{BAO5,BAO6} as  $\Delta z^{obs}$ and   $(1+z)\Delta \theta^{obs}$   on Table \ref{dzobs}, a manner convenient for this work. In spherically symmetric space-times $1/\Delta z=\alpha_{\parallel} H^{fid} r_d^{fid}/c$ and   $[(1+z)\Delta \theta^{obs}]^{-1}=\alpha_{\perp}D_A^{fid}/r_d^{fid}$, where superscript $fid$ refers to a homogeneous fiducial model,  $r_d$ to a comoving sound horizon at the drag epoch and $\alpha_{\parallel}=\Delta z^{fid}/\Delta z$ and $\alpha_{\perp}=\Delta \theta^{fid}/\Delta \theta$. In homogeneous space-times also  the relations $1/\Delta z=H r_d/c$ and   $[(1+z)\Delta \theta^{obs}]^{-1}=D_A/r_d$ hold, hence reporting the results as $H r_d/c$ and $D_A/r_d$, as is done in Refs.  \cite{BAO5,BAO6}, is merely a homogeneous interpretation of the results.    

We assume no correlation between the high and the low redshift observations, hence  the combined $\chi^2$ for the low and a given \lya  observation in the radial direction is 
\be{chitotR}
(\chi_{low+L\alpha FR})^2=(\chi_{low })^2+\frac{(\Delta z^{obs}-\Delta z)^2}{\sigma^2_{L\alpha FR}},
\ee
and in the transverse direction is
\be{chitotT}
(\chi_{low+L\alpha FT})^2=(\chi_{low })^2+\frac{[(1+z)\Delta \theta^{obs}-(1+z)\Delta \theta]^2}{\sigma^2_{L\alpha FT}}.
\ee
Here
\bea{chicombLTBR}
\Delta z^{LTB}&=&(1+z_{L\alpha F})H^R_{L\alpha F}\frac{R_{,r}^{L\alpha F}}{R_{,r}(r_{L\alpha F},t_e)} l_{phys}(t_e) , \\ \Delta z^{F}&=&(1+z_{e})H_{L\alpha F} l_{phys}(t_e), \\
\Delta \theta^{LTB}&=&  (1+z_{L\alpha F})H^R_{L\alpha F}\frac{R_{,r}^{L\alpha F}}{R_{,r}(r_{L\alpha F},t_e)} l_{phys}(t_e), \\
\Delta \theta^{F}&=&(1+z_{e})H_{L\alpha F} l_{phys}(t_e),
\eea
and the observed values are tabulated on Table \ref{dzobs}. We study the following five cases separately:
\begin{itemize}
\item[-] low BAO
\item[-] low BAO and L$\alpha$F auto-correlation in the radial direction (L$\alpha$FacR)
\item[-] low BAO and L$\alpha$F auto-correlation in the transverse direction (L$\alpha$FacT)
\item[-]   low BAO and L$\alpha$F cross-correlation in the radial direction (Ly$\alpha$FccR) 
\item[-]  low BAO and L$\alpha$F cross-correlation in the transverse direction (Ly$\alpha$FccT)
\end{itemize}
These five different cases are investigated separately because of the acknowledged  discrepancy between L$\alpha$F data and the \lcdm model, which has raised  suspicions of  the validity of the measurements (see \cite{Planck2015}). We expect these  cases to reveal if discrepancy exists also in the void models. Moreover, we also expect to  obtain some comprehension  of the effects of the systematic errors. 

 \begin{table}
\centering
  \begin{tabular}{| p{1.8cm} | c | c | c | c | c | c | c |}
    \hline
    data sets   & model &  $\chi_{min}^2$ & $\Omega_{m,min}^F$ & $\sigma$ & $2\sigma$ & $3\sigma$ & $1-p$ \\ \hline 
 \multirow{2}{*}{  low BAO} & \lcdm   & 1.7  & 0.23 & $[0.11 , 0.39]$  & $[0.075,0.47]$ & $[0.048,0.54]$ & 0.21 \\
  &LTB  &  2.6  & 0.47 & $[0.38,0.58]$ & $ [0.36,0.64]$ & $[0.34,0.69]$ & 0.38 \\ \hline
    \centering{ low BAO} + L$\alpha$FacR  & \parbox[t]{1.1cm}{ \lcdm \\ \centering{LTB}} & \parbox[t]{.5cm}{1.7\\15} & \parbox[t]{.67cm}{0.23\\0.35}&   \parbox[t]{1.75cm}{$[0.19,0.27]$ \\$[0.31,0.39]$} & \parbox[t]{1.75cm}{$[0.18,0.29]$\\$[0.30,0.41]$} & \parbox[t]{1.75cm}{$[0.17,0.31]$\\$[0.30,0.42]$} &  \parbox[t]{0.67cm}{0.11\\0.99} \\ \hline
    \centering{ low BAO} + L$\alpha$FccR  & \parbox[t]{1.1cm}{ \lcdm\\ \centering{LTB}} &  \parbox[t]{.5cm}{1.7\\13} & \parbox[t]{.67cm}{0.24\\0.36} & \parbox[t]{1.75cm}{$[0.20,0.28]$ \\$[0.32,0.41]$} & \parbox[t]{1.75cm}{$[0.18,0.30]$\\$[0.31,0.42 ]$} & \parbox[t]{1.75cm}{$[0.17,0.32]$\\$[0.30,0.44 ]$} &\parbox[t]{0.67cm}{0.11\\0.98} \\ \hline
    \centering{ low BAO} + L$\alpha$FacT  & \parbox[t]{1.1cm}{ \lcdm \\ \centering{LTB}} & \parbox[t]{.5cm}{2.1\\2.9} & \parbox[t]{.67cm}{0.26\\0.48}&   \parbox[t]{1.75cm}{$[0.16,0.40]$ \\$[0.40,0.59]$} & \parbox[t]{1.75cm}{$[0.13,0.48]$\\$[0.38,0.65]$} & \parbox[t]{1.75cm}{$[0.10,0.54]$\\$[0.36,0.69]$} &  \parbox[t]{0.67cm}{0.16 \\ 0.28} \\ \hline
    \centering{ low BAO} + L$\alpha$FccT  & \parbox[t]{1.1cm}{ \lcdm\\ \centering{LTB}} &  \parbox[t]{.5cm}{4.0\\4.8} & \parbox[t]{.67cm}{0.33\\0.53} & \parbox[t]{1.75cm}{$[0.23,0.47]$ \\$[0.44,0.64]$} & \parbox[t]{1.75cm}{$[0.19,0.53]$\\$[0.42,0.69]$} & \parbox[t]{1.75cm}{$[0.17,0.59]$\\$[0.40,0.73]$} &  \parbox[t]{0.67cm}{0.45 \\0.56} \\ \hline
  \end{tabular}\caption{\footnotesize The minima of $\chi^2$ and corresponding $\Omega_m^F$ for different
data sets and models are tabulated. The minima appear to be independent
on $H_0^F$ and the $n \sigma$-limits are given for $\Omega_m^F$. The
$p$-values are given as $1-p$ on the last column.  } 
 \label{t2} \end{table}

\begin{figure}[t]
\centering
\includegraphics[scale=0.88]{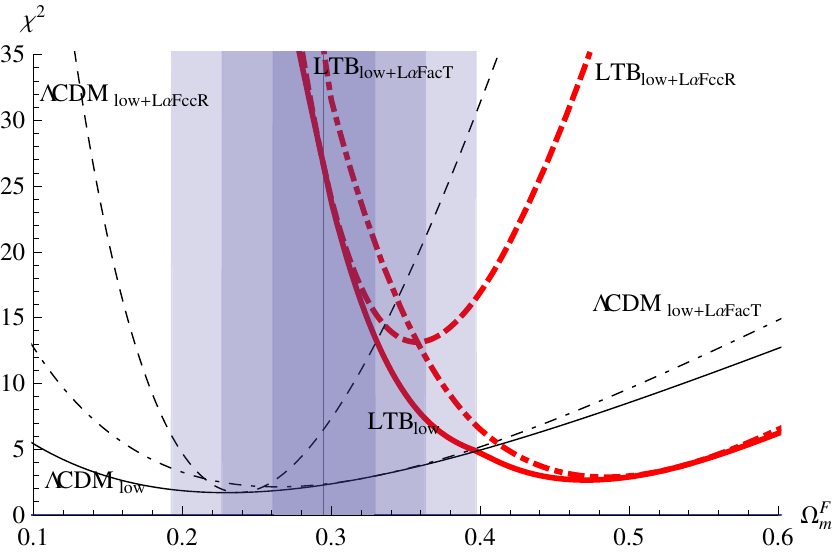} \, \,
\includegraphics[scale=0.88]{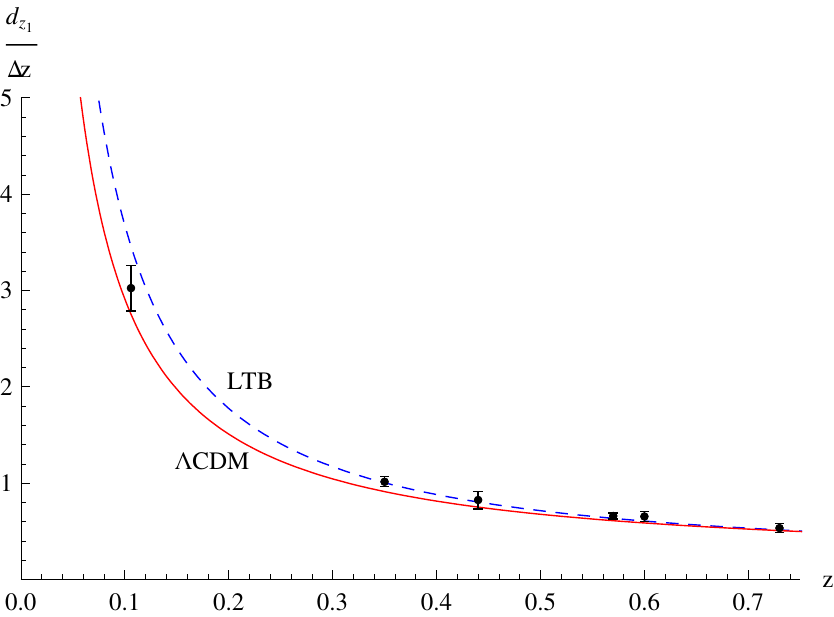} 
\caption{\footnotesize The left panel: Different $\chi^2(\Omega_m^F)$ curves for different models and data sets are presented. The black vertical line in the middle of the vertical contours indicates  the preferred $\Omega_m^F$ according to the SN data \cite{SN}, whereas the vertical contours indicates 1-3 $\sigma$ deviations from the best fit value. See Table \ref{t2} for numerical values. The right panel:  The (black) points and bars represent observations and their $1 \sigma$ errors, the red solid curve represents the \lcdm model's prediction and the dashed blue curve represent the LTB model's prediction. For both models, $\Omega_m^F=0.32$ and $\Delta z$ is chosen to correspond L$\alpha$FccR. Colors available online.}
\label{contour}
\end{figure}

 The analysis is executed so that for each $\Omega_m^F$ and $H_0^F$ we find the minimum $\chi^2_{min}$ for both models by allowing $l_{phys}$ to vary. We constrain parameters $\Omega_m^F$, $H_0^F$  and $l_{phys}(t_e)$ so that \footnote{We were unable to execute the procedure described in Section \ref{results} for  $\Omega_m^F< 0.2$. Therefore, the $\chi^2$ was not calculated for  the LTB model for these $\Omega_m^F$ values. We identified the reason to be that  $\Omega_m^F< 0.2$ requires $\Omega_M(0)<0$. }
\be{range}
 0.01\leq \Omega_m^F \leq 0.85, \qquad  30\leq H_0^F \leq 80, \qquad  0\leq  l_{phys}(t_e).
\ee
We find the $\chi^2$ to be independent of $H_0^F$ for both models. Numerical analysis reveals that $d_{z}^{LTB}$, $\Delta z^{LTB}$, and $\Delta \theta^{LTB}$ depend only on the combination  $(H_0^F)^{1/3} l_{phys}(t_e)$, whereas $d_{z}^{F}\propto \Delta z^{F} \propto \Delta \theta^{F} \propto H_0^F l_{phys}(t_e)$ [see Eq.\ \eqref{dzf2}]. Consequently, we can fix $H_0^F$ using local Hubble observations solely, thus leaving us two free parameters to fit the BAO observations. The curves $\chi_{low}^2(\Omega_m^F)$ for both of the models are plotted in Figure \ref{contour}, together with  the latest SN constraints presented in \cite{SN}.\footnote{We use supernova constraints here for two reasons. We took the  \lcdm  luminosity distance for the LTB model for the very reason the models fit equally well to the supernova observations. In addition, it is not evident how the Planck results \cite{Plank} should be interpreted here. As they are obtained from the CMB sky,  their feasibility  constraining  the inhomogeneous models that fits supernovae is questionable.  } Both models can explain the used low BAO data with comparable minimum $\chi^2_{low}$ (see Table \ref{t2}) though the LTB model is  in distinct conflict with the SN data. Also  $\chi_{low+L\alpha FccR}^2(\Omega_m^F)$ and $\chi_{low+L\alpha FacT}^2(\Omega_m^F)$ curves are drawn in Figure \ref{contour}.  These specific $\chi_{low+Ly\alpha F}^2(\Omega_m^F)$ curves were chosen, because they lay less stringent constraints to the models in radial and transverse directions compared with the alternatives (see Table \ref{t2}).

For any of the five  combinations of low and high BAO studied above, we can include the SN consraints \cite{SN} and  calculate  $\chi_{tot}^2$ by
\be{chisqrdtot}
\chi_{tot}^2(\Omega_m^F)=\chi_{BAO}^2(\Omega_m^F)+\frac{(\Omega_m^F-\Omega_m^{obs})^2}{\sigma_{obs}^2},
\ee
where we use  $\Omega_m^{obs}=0.295\pm 0.034$ obtained from \cite{SN}. The results are presented on Table \ref{t3}.

We use Bayesian information criteria (BIC) and $p$-value to asses the strength of the models. Because both models have equal amount free parameters and the used errors are Gaussian, the difference in the BIC values of the models is 
\be{BIC}
\Delta BIC = \left[\chi^{LTB}_{min}(\Omega_m^F)\right]^2-\left[\chi^F_{min}(\Omega_m^F)\right]^2.
\ee
A difference in BIC of 2 is considered positive evidence against the model with the higher BIC, in this case higher $\chi^2$ corresponds to higher BIC, 
while a difference in BIC of 6 (or more) is considered strong evidence \cite{Liddle2004}. For further information, we refer to \cite{Davis2007} and \cite{Schwarts1978}.

All the evidence  obtained here using the maximum likelihood analysis favours the \lcdm model.  Especially for the combined SN and BAO data, the $\Delta$BIC values exhibit strong evidence against the LTB model and $p$-values rule out the LTB model at least at $98 \%$ confidence (see Table \ref{t3}).  Figure \ref{contour} illustrates the effect of \lya data for both models. The discrepancy between SN, low BAO and high BAO is present for both models. For the LTB model, the transversal \lya data appears to be consistent with  low BAO, but in severe conflict with SN data, whereas the radial \lya data appears not to conflict much with SN data, but is in a clear contradiction with low BAO observations. The weaker constraining curves drawn in Figure \ref{contour} demonstrate the situation well on a qualitative level and using the stringent constraining curves would not modify these conclusions. Therefore, we conclude the systematic errors in \lya observations do not play a significant role here.

 \begin{table} \label{t3}
\centering
  \begin{tabular}{| p{1.55cm} | c | c | c | c | c | c | c | c |}
    \hline 
    data sets   & model &  $\chi_{min}^2$ & $\Omega_{m,min}^F$ & $\sigma$ & $2\sigma$ & $3\sigma$ & $1-p$ & $\Delta$BIC \\ \hline
   \parbox[t]{1.55cm}{\vspace{0.001cm} low$\, + \,$SN}&\parbox[t]{1.1cm}{ \lcdm  \\ \centering{LTB}}  &  \parbox[t]{.4cm}{2.2\\12}  & \parbox[t]{.6cm}{0.29\\0.37} &  \parbox[t]{1.71cm}{$[0.23,0.34]$\\$[0.33,0.40]$} &\parbox[t]{1.71cm}{$[0.21,0.36]$\\$[0.32,0.42 ]$} &\parbox[t]{1.71cm}{$[0.19,0.38]$\\$[0.31,0.44]$}  &  \parbox[t]{.6cm}{0.30\\0.98} &  \parbox[t]{0.4cm}{\vspace{0.001cm}10}\\ \hline
    low$\, + \,$SN $+ \,$L$\alpha$FacR  & \parbox[t]{1.1cm}{ \lcdm \\ \centering{LTB}} & \parbox[t]{.4cm}{4.2\\17} & \parbox[t]{.6cm}{0.25\\0.34}& \parbox[t]{1.71cm}{$[0.22,0.29]$\\$[0.31,0.37]$} &\parbox[t]{1.71cm}{$[0.20,0.30]$\\$[0.30,0.38]$} &\parbox[t]{1.71cm}{$[0.19,0.31]$\\$[0.29,0.39]$} &  \parbox[t]{.6cm}{0.48\\1.00} &  \parbox[t]{0.4cm}{\vspace{0.001cm}13} \\ \hline
    low$\, + \,$SN $\, + \,$L$\alpha$FccR  & \parbox[t]{1.1cm}{ \lcdm\\ \centering{LTB}} &  \parbox[t]{.4cm}{3.5\\16} & \parbox[t]{.6cm}{0.26\\0.34} & \parbox[t]{1.71cm}{$[0.22,0.30]$\\$[0.32,0.37]$} &\parbox[t]{1.71cm}{$[0.21,0.31]$\\$[0.31,0.39]$} &\parbox[t]{1.71cm}{$[0.20,0.33]$\\$[0.30,0.40]$} &  \parbox[t]{0.6cm}{0.38\\0.99} &   \parbox[t]{0.4cm}{\vspace{0.001cm}12}\\ \hline
    low$\, + \,$SN $+ \,$L$\alpha$FacT  & \parbox[t]{1.1cm}{ \lcdm \\ \centering{LTB}} & \parbox[t]{.4cm}{2.3\\15} & \parbox[t]{.6cm}{0.29\\0.38}& \parbox[t]{1.71cm}{$[0.23,0.34]$\\$[0.34,0.42]$} &\parbox[t]{1.71cm}{$[0.21,0.37]$\\$[0.33,0.44]$} &\parbox[t]{1.71cm}{$[0.20,0.39]$\\$[0.32,0.45]$} &  \parbox[t]{.6cm}{0.19\\0.99} &  \parbox[t]{0.4cm}{\vspace{0.001cm}13} \\ \hline
    low$\, + \,$SN $\, + \,$L$\alpha$FccT  & \parbox[t]{1.1cm}{ \lcdm\\ \centering{LTB}} &  \parbox[t]{.4cm}{4.2\\23} & \parbox[t]{.6cm}{0.30\\0.40} & \parbox[t]{1.71cm}{$[0.25,0.36]$\\$[0.36,0.44]$} &\parbox[t]{1.71cm}{$[0.23,0.38]$\\$[0.35,0.46]$} &\parbox[t]{1.71cm}{$[0.22,0.40]$\\$[0.34,0.47]$} &  \parbox[t]{0.6cm}{0.48\\1.00} &   \parbox[t]{0.4cm}{\vspace{0.001cm}19}\\ \hline
  \end{tabular}\caption{\footnotesize  The minima of $\chi^2$ and corresponding $\Omega_m^F$ for different
data sets and models are tabulated. The minima appear to be independent
on $H_0^F$ and the $n \sigma$-limits are given for $\Omega_m^F$. The
$p$-value and $\Delta$BIC are given on the two last column, the
$p$-value reported as $1-p$ . The SN data is obtained from Ref. \cite{SN}.} 
\label{t3} \end{table}

\subsubsection{Scale independent analysis}

In addition to the $\chi^2$ analysis, we compare the models using a method independent on $l_{phys}$. For this analysis, we choose to use L$\alpha$F data  for which the $\chi^2$ values of the models are comparable and $\Omega_m^F$ value acceptable in the light of SN data. For these reasons, we choose to use low BAO and  L$\alpha$FccR data and  fix $\Omega_m^F=0.32$. 
Taking the  ratio $d_z/\Delta z$ eliminates the dependence on $l_{phys}$ and at redshifts $z_1$ and $z_2$ we find
\be{dzLT2ratio}
\frac{d_{z_1}^{LTB}}{\Delta z_2^{LTB}}=\left(\frac{1+z_{1}}{(D_A^{1})^2} \frac{H^R_{1}}{z_{1}}\right)^{1/3}\left[(1+z_{2}) H^R_{2}\right]^{-1}\left(\frac{\mathcal{R}^R_1}{\mathcal{R}^R_2}\right)^{1/3}\left(\frac{\mathcal{R}^T_1}{\mathcal{R}^R_2}\right)^{2/3}.
\ee
 The \lcdm counterpart   of the above equation is
\be{dzF2ratio}
\frac{d_{z_1}^{F}}{\Delta z_2^{F}}=\left(\frac{1+z_{1}}{(D_A^{1})^2} \frac{H^F_{1}}{z_{1}}\right)^{1/3}\left[(1+z_{2}) H^F_{2}\right]^{-1}\frac{\mathcal{R}^F_1}{\mathcal{R}^F_2}.
\ee
The difference between Eqs.\  \eqref{dzLT2ratio} and \eqref{dzF2ratio} is caused by  the Hubble functions and the relative local expansion rates of the different models.   The ratios $d_{z_1}/\Delta z_2$, where  $\Delta z_2=\Delta z_{L\alpha FccR}$, are plotted in  Figure \ref{contour}. The models appear to fit equally well to the observations, but calculating  the variance,
\be{var}
Var=\sum_{z_1}\left( \frac{d_{z_1}}{\Delta z_{2}}- \frac{d^{obs}_{z_1}}{\Delta z_{2}^{obs}}\right)^2,
\ee
which for LTB is   $\approx 0.11$  and for \lcdm is $\approx 0.19$, reveals that the LTB model gives slightly a better fit.
This is in minor disagreement with the maximum likelihood analysis, which gives  a slightly lower $\chi^2$ value for the \lcdm than for the LTB, as can be seen from Figure \ref{contour}. The small disagreement can be understood by considering the nature of the ratio analysis. Because the variance does not take standard deviations into account, $\Omega_m^F$ is fixed and $l_{phys}$ is eliminated (unlike in the maximum likelihood analysis where they were optimised), is variance \eqref{var} merely a measure how much data points differ from  fixed models. From Table \ref{t2} can be seen that the difference between $\Omega_{m,min}^F$ of low BAO + L$\alpha$FccR and the fixed $\Omega_m=0.32$ is 0.4 for the LTB and 0.8 for the \lcdm model. Even though the values in the table correspond to maximum likelihood analysis, we expect the corresponding $\Omega_{m,min}^F$ values for the variance to be in the vicinity, thus explaining the small difference in the  results of two  analysis methods.

The ratio analysis demonstrates how the differences between the Hubble functions and RLER's of the two models are inherited by the measurable BAO quantities. Figure \ref{contour} demonstrates how the difference in the predictions of the BAO observables of the models increases with respect to redshift, as expected (see Figure \ref{k11}).\footnote{Note that the curves $d_{z_1}/\Delta z$ of the models do not converge when $z$ grows larger than presented in Figure \ref{contour}. Instead, they cross at $z\approx 0.85$ and   $d_{z_1}/\Delta z$ is 0.27 for the \lcdm and 0.26 for the LTB at $z=2.34$.}    At the same time, the errors appear to grow with redshift, hence the constraining power of the wide range of the observations is not strong with this method.

We deduce the results obtained here  are mainly caused by the RLER's for two reasons. The RLER's differ more than  the Hubble functions   at the redshift range of the observations.  Moreover, $d_{z_1}/\Delta z_2$ in Eqs.\ \eqref{dzLT2ratio} and \eqref{dzF2ratio} is effectively  proportional to the power of $-2/3$ of the Hubble functions, whereas they are effectively linearly dependent on RLER's.

Both of the two analyses showed that BAO observables do inherit the growing constraining power with respect to redshift from the Hubble functions and LRER's. Minimum likelihood analysis clearly favours the \lcdm model, but demonstrated the models to be comparable if \lya data in the radial direction is employed and $\Omega_m^F$ is close to the value favoured by Planck \cite{Planck2015}. Analysing this specific case with a scale independent method, we found the LTB model to be slightly favoured. However, it is not evident how homogeneously interpreted Plank results should be received here. Moreover, we expect that the scale independent analysis using transverse \lya data would yield a contrary result.    On the strength of these results, we conclude that the \lcdm model fits combined BAO and SN observations better than the LTB model.

\subsection{Cosmic microwave background} \label{CMB}

The angular diameter of the decoupling surface  $\theta^*$ is in the LTB model given by 
\be{theta}
\theta^*= \frac{l^*_{phys}(r^*,t^*)}{R(r^*,t^*)}.
\ee
 In practical calculations, we take the decoupling time $t^*$ to correspond  to a redshift $z=1089$. The latest Planck results \cite{Plank} give $\theta^*=(1.04147 \pm0.00062)\times 10^{-2}$. Since at early times, the mean physical lengths are assumed to be the same at every $r$, Eq.\ \eqref{dzLT2} can also be written 
\bea{dzLT3}
d_{z_{BAO}}^{LTB}&=&\left(\frac{1+z_{BAO}}{(D_A^{BAO})^2} \frac{H^R_{BAO}}{z_{BAO}}\right)^{1/3}\left(\frac{\mathcal{R}^R_{BAO}}{R_{,r}(0,t_d)}\right)^{1/3}\left(\frac{\mathcal{R}^T_{BAO}}{R(0,t_d)}\right)^{2/3} l_{phys}(r_d,t_d) ,
\eea
where the index $d$ refers to drag epoch, for which we use the redshift $z= 1020$. At the homogeneous era
\be{ratio}
 \frac{l_{phys}(r^d,t^d)}{l_{phys}(r^*,t^*)}= \frac{R_{,r}(r^d,t^d)\int_{r^d}^{r^d+r^d_s}dr}{R_{,r}(r^*,t^*)\int_{r^*}^{r^*+r^*_s}dr}= \frac{a(t^d)\int_{r^d}^{r^d+r^d_s}dr}{a(t^*)\int_{r^*}^{r^*+r^*_s}dr}=\frac{1+z^*}{1+z^d} \frac{r^d_{s}}{r^*_{s}}=\frac{1090}{1021}( 1.0188 \pm 0.0068 ),
\ee where the numerical value for $r_s^d/r_s^*$ is taken from Ref. \cite{Planck2015}. Combining \eqref{theta}-\eqref{ratio}, we can write
\bea{dzltpertheta}
\frac{d_{z_{BAO}}^{LTB}}{\theta^*}
&=&\left(\frac{1+z_{BAO}}{(D_A^{BAO})^2} \frac{H^R_{BAO}}{z_{BAO}}\right)^{1/3}\left(\frac{\mathcal{R}^R_{BAO}}{\mathcal{R}^R_{d}}\right)^{1/3}\left(\frac{\mathcal{R}^T_{BAO}}{\mathcal{R}^T_{d}}\right)^{2/3}\frac{1+z^*}{1+z^d} \frac{r^d_{s}}{r^*_{s}}R(r^*,t^*),
\eea
since $R_{,r}(0,t_d)=A(0,t_d)=\mathcal{R}^T_{d}=\mathcal{R}^R_{d}$. The \lcdm counterpart equation is
\bea{dzfpertheta}
\frac{d_{z_{BAO}}^{F}}{\theta^*}
&=&\left(\frac{1+z_{BAO}}{(D_A^{BAO})^2} \frac{H^F_{BAO}}{z_{BAO}}\right)^{1/3}\mathcal{R}^F_{BAO}(1+z^*) \frac{r^d_{s}}{r^*_{s}}D_A^*.
\eea

\begin{figure}[t]
\centering
\includegraphics[scale=0.9]{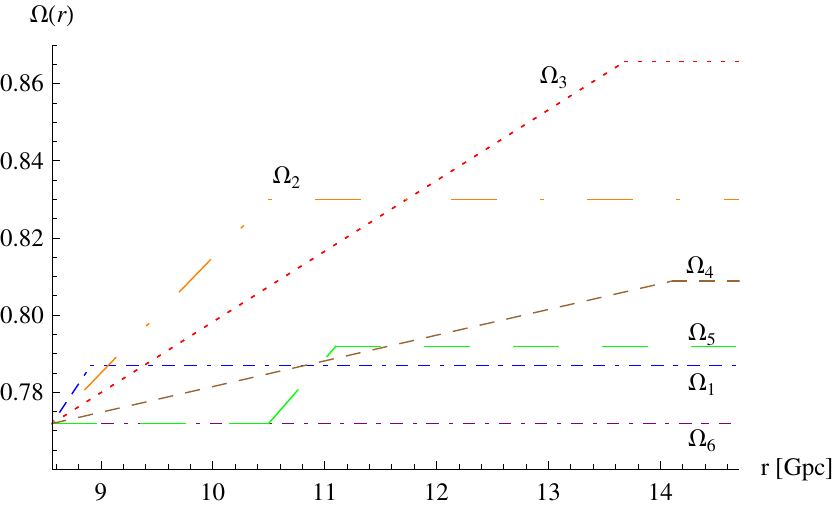} 
\includegraphics[scale=0.9]{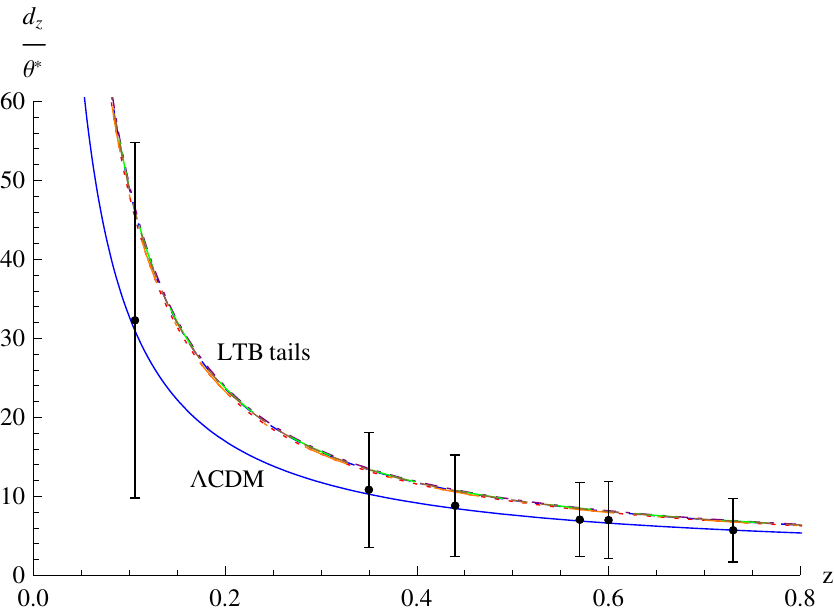} 
\caption{\footnotesize  The LTB and the  \lcdm models with $H_0^F=67.1$ and $\Omega_m^F=0.32$. In the left panel several tails are plotted. In the right panel, (black) points and bars are observed $d_z/\theta^*$ values and their $1\sigma$ error bars, the solid blue curve represents the \lcdm prediction and non-solid colored curves represent the LTB models with tails, the curve color and style indicates the correspondning left panel tails. The LTB curves can be arranged from the worst to the best fit as follows: $\Omega_6$,   $\Omega_5$,   $\Omega_1$,  $\Omega_4$,    $\Omega_2$,  and   $\Omega_3$. Colors available online. }
\label{bk2b}
\end{figure}

Eqs.\   \eqref{dzltpertheta} and \eqref{dzfpertheta} include quantities dependent on the BAO, the drag and the decoupling epochs. By choosing how to continue the void profile function beyond the SC singularity (the tail) one fixes the  values of the quantities dependent on  the drag and the decoupling epochs. On the other hand, we have fixed $\Delta z_2=\Delta z_{L\alpha FccR}$.  Therefore, $d_{z_1}^{LTB}/\Delta z_{2}^{LTB} \propto d_{z_{BAO}}^{LTB}/\theta^*$ and  $d_{z_1}^{F}/\Delta z_{2}^{F} \propto d_{z_{BAO}}^F/\theta^*$.   Eqs.\   \eqref{dzltpertheta} and \eqref{dzfpertheta} predict the BAO observables  as rescalings of  Eqs.\   \eqref{dzLT2ratio} and \eqref{dzF2ratio}. This is illustrated in   Figures \ref{contour} and  \ref{bk2b}. This implies that regardless how the void profile is continued beyond the SC, the LTB model can hardly describe the wide range of BAO observables better than in the case where only BAO was investigated.

Despite of the rescaling effect noted above, it is not evident that optimal tails are natural for the LTB model. Therefore, we consider several tails (illustrated in Figure \ref{bk2b}) to be able to compare the models.  The tails are chosen requiring that the SC condition is not violated and that the system  is approximately homogeneous at $z\gtrsim 1000$. The latter condition is satisfied by constant tails for $z\geq 1020$. However, this condition does not guarantee the universe to be homogeneous  at $z\simeq 1020$, but merely enables it (see Section \ref{results}). The tails in the left panel of Figure \ref{bk2b} are constructed as follows: The blue tail $\Omega_1$ is rising rapidly and changing to a constant before the SC singularity is violated. The orange  curve $\Omega_2$ is tailored to rise slightly slower than the blue curve and changes to a constant before the SC singularity occurs. The red  tail $\Omega_3$ is rising to encounter the SC singularity and the homogeneity limit almost at the same time. The brown  tail   $\Omega_4$ rises slowly before it reaches the homogeneity limit. The green  tail $\Omega_5$ is set to be a constant at first, ascent after a while and turn to a constant again before the SC condition is violated. The purple  curve  $\Omega_6$ is set to a constant from $z\approx 6.9$.  

In the right panel of Figure \ref{bk2b}  we have  plotted the  LTB and the \lcdm predictions for the ratio $d_z/\theta^*$ using different void profiles together with the observed values. The models in question are characterised by $H_0^F=67.1$ and  $\Omega_m^F=0.32$.  As can be seen from the figure, changing the void tail does not seem to have great impact on the predicted ratios. Even though the LTB models fit well inside the error bars, they are characterised by the feature that they do not fit  observations as well as the \lcdm model.     The LTB models can be arranged from the best fit to the worst fit. It appears that higher $\Omega_{M}$ maximum  value and a steeper incline gives a better fit to the data. The problem of the linear piecewise functions is that they can not be as steep and rise as high as needed without violating either the shell crossing condition \eqref{sc} or the homogeneity assumption.

We also applied more complex functions to describe the tail and considered the possibility of it starting  at smaller redshifts. Evidently, this causes the luminosity distance of the LTB model to begin to deviate from the \lcdm model at lower redshifts.  Although tails with a steep ascent to high maximum values made the fit better, they did not differ considerably from the piecewise tails.

 The difference between Eqs.\  \eqref{dzltpertheta} and \eqref{dzfpertheta} is caused by the  Hubble functions, the RLER's  and the angular diameter distances after the SC. Although both of the equations include the factor $r^d_s/r^*_s $, a somewhat reasonable assumption is that it does not take the same value in both models. After all, $r^d_s/r^*_s \simeq 1$ implies assuming  the baryon fraction is the same at the BAO, drag and decoupling epochs. With a variable baryon fraction one would obtain
\be{ratio2}
\frac{r^d_s}{r^*_s}=\tilde{c} \,(1.0188 \pm 0.0068 ),
\ee 
where  $\tilde{c}$ is a constant,  determined by the amounts of baryonic and dark matter at $r_d$ and $r_{BAO}$. One can choose this constant to give the best possible fit for the LTB model, which corresponds to $\tilde{c}\approx 0.7$. Nevertheless, as we showed earlier,  $d_{z_1}^{LTB}/\Delta z_{2}^{LTB}$ and  $d_{z_1}^{F}/\Delta z_{2}^{F}$ are just rescaled  $ d_{z_{BAO}}^{LTB}/\theta^*$ and  $ d_{z_{BAO}}^F/\theta^*$, respectively, thus  the combined BAO and CMB set is at the best at the same level as the fit obtained using BAO observations only. Therefore, we conclude that the inclusion of CMB observations  emphasizes the superiority of the \lcdm model over the LTB model.

Note that the LTB models are dust only solutions of the Einstein equations. Consequently, the effects of radiation are neglected here and for the reason the treatment of the  CMB is incomplete. However, our analysis starts at $z=1089$, when the radiation density was still subdominant while not completely negligible. Therefore, our studies of the CMB are pointing  how difficult it is for the LTB void to mimic different features of the \lcdm model at wide redshift range. Moreover, because the \lcdm model is in good concordance with the first CMB peak,  we are confident our conclusion of the inclusion of the CMB data holds even if the effects of pressure would be considered.   For discussion about the dynamical effects of radiation in inhomogeneous models, see \cite{ClarksonRegis2011}. 

\section{Conclusions and discussion} \label{conclusions}

We have compared the  \lcdm and the LTB models which have the same distance - redshift relation and local Hubble parameter value. In addition, the LTB model is forced to have a homogeneous bang time  and to be  homogeneous between the bang time and the drag epoch. The relative local expansion rates appeared to have the largest difference, but the Hubble functions showed considerable deviation too. The difference between the \lcdm and the LTB models is larger the wider the redshift range covered. On the strength of this, we investigated  baryon acoustic oscillation and cosmic microwave background observations. The results  favour the \lcdm model. Furthermore, the LTB models  exhibits conflict between SN and BAO observations (with or without \lya data).  They suggest, however, that  to obtain comparable fits for SNIa, $H_0$, CMB, and BAO, one needs to be able to control the LTB model's luminosity distance, Hubble functions, and relative local expansion rates. To achieve this, the model needs  more degrees of freedom than those employed in this paper.

The analysis of low redshift BAO observations slightly  favoured the \lcdm to the LTB model. Using the maximum likelihood analysis by allowing  $\Omega_m^F$ and $H_0^F$ to vary, we noticed that  the minimum $\chi^2$ value is independent on $H_0^F$ for both models. For low  BAO observations  $\chi^2_{min}$ for the \lcdm model is 1.7 and for the LTB model is 2.6.    Even though these values are close to each other, for the LTB model it was obtained when the mimicked \lcdm model has $\Omega_m^F=0.47$, which is in conflict with the latest SN \cite{SN} and Planck \cite{Plank} data. Thus, although both models can comparably explain low  BAO observations, the combined low BAO and SN data clearly favours the \lcdm model. This deduction is endorsed by $p$-value, which states the LTB model to be ruled out  by $98 \%$ confidence and Bayesian information criteria stating there is strong evidence against the LTB model.

We also investigated the effects of the baryonic features of the Lyman-$\alpha$ forest. For the combined set of low redshift BAO and L$\alpha$F data in the radial direction, the LTB model predicts the  set   distinctly worse than low BAO only, but the tension to SN  does not grow  appreciably. The effect on the \lcdm model is reverse as adding the radial \lya observations have almost negligible effect on explaining the  BAO,  but tension to SN grows drastically. For the combined BAO and supernova data $p$ and  BIC values clearly favour the \lcdm model, even though the results are not complementing it either. The weakest constraint from $p$-values rule the \lcdm out by $38 \%$ confidence and the LTB out by $98 \%$ confidence. Moreover, the BIC value indicates  strong evidence against the LTB model.  Furthermore, if  low BAO data is combined with the transverse \lya data, the results are different from the radial case. For the LTB model, the transversal \lya data appears to be consistent with low BAO, but in severe conflict with SN data, whereas these  three data sets flatter the \lcdm model. The weakest constraint from $p$-values rule the \lcdm out by $19 \%$ confidence and the LTB out by $99 \%$ confidence and   the BIC value indicates  strong evidence against the LTB model. In conclusion,   the \lcdm model fits  BAO (with or without \lya data) and SN observations better than the LTB model.

In addition, we compared how LTB   and \lcdm compare to BAO and first CMB peak observations. The LTB model can not mimic the \lcdm luminosity distance -  redshift relation at the shell crossing singularity, of which location depends on the parameters of the mimicked \lcdm model; for parameters $H_0^F=67.1$ and $\Omega_m^F=0.32$ the location is at $z\approx 6.9$.  For this reason, we modified the void profile  beyond $z\approx 6.9$ for this particular model,  and compared it with observations. We studied different tails, but no significant improvement was found.  Only if the mean sound horizon depends on the position, some remedy was achieved. This possibility is related to the assumption of inhomogeneous early time, indicating our assumption of its homogeneity is too restrictive. Another deficiency in our approach is that the LTB model does not include pressure, thus leaving our CMB analysis incomplete. For these reasons, we conclude that under the assumptions made in this study, combining CMB  to the BAO observations emphasises the better fit of the \lcdm model. However, due to the concordance of the \lcdm model and the CMB observations, we expect this to hold quite generally.

Explaining the local Hubble value observations does not cause difficulties to the LTB model studied here, unlike in some other studies. The reason is our parametrisation, which allows us to satisfy the $H_0$ observations independently on other observations. This is not the case e.g.\ in \cite{BNV2010}, where the other observables fix all free parameters and $H_0$ is then to be calculated. In \cite{ZG-BR-L2012} the parameterisation allows freely to choose $H_0$ for the model, but the authors find observably viable values to correspond to too young universe. This suggests that the issues confronted in explaining $H_0$ observations could vanish for some LTB models by reparametrisation.

Our results agree with studies \cite{ZG-BR-L2012,MossZibinScott2011,RedlichEtAl2014}, but appears to be in conflict with \cite{Garcia-BellidoHaugbolle2008a} and \cite{BNV2010}.
The LTB void models were found to have comparable fit to the \lcdm model using  SNIa, first CMB peak and BAO observations in  \cite{Garcia-BellidoHaugbolle2008a} and in addition to $H_0$ observations in \cite{BNV2010}, even though in the latter case, difficulties creating high enough $H_0$ value were  encountered.  We identify the factors explaining these differences.  In these articles, the sound horizon was implicitly allowed to vary with distances.  Moreover, in both papers, the redshift range of the BAO observations was narrower even than the low BAO data in this article. We demonstrated how the deviation of the BAO features between the two models grows with respect to redshift. Because the trusted low BAO observations are in concordance with the \lcdm model, it appears to be justified to say that the wider redshift range imply worse concordance with the LTB model and low BAO observations. This is backed by Ref. \cite{ZG-BR-L2012}, where as wide redshift range of BAO observations was utilized as with low BAO here and the fit between observations and the models was far worse. 

We believe that we have found the reasons why LTB models constantly fails to explain $H_0$, SNIa, BAO and CMB observations simultaneously. Combining SNIa observations, that determine the luminosity distance - redshift relation, with BAO observations, which are heavily dependent on the  local expansion rates, tension is manifested between the model and the observations. Extending the range from where the observations dependent on the  local expansion rate and the Hubble function are obtained, such as including CMB and local Hubble parameter in the range, the tension grows.  This suggests that  to obtain observationally more viable models, one needs to include more degrees of freedom  by renouncing the homogeneous early universe hypothesis,  the homogeneity of the bang time, the negligible effects of pressure or the spherical symmetry assumption.

\acknowledgments

This study is partially (P. S.) supported by the Magnus Ehrnroothin S\2\2ti\3 foundation (6.3.2014) and Turun Yliopistos\2\2ti\3 foundation, identification number 11706. E.M. acknowledges support from the Swedish Research Council.

%%%%%%%%%%%%%%%%%%%%%%%%%%%%%%%%%

\appendix
\section{Enhancing the accuracy of the solutions} \label{AA}

To increase the numerical accuracy of the solutions of the system, we modify the system of differential equations  \eqref{rs}, \eqref{ng}, \eqref{add2} and \eqref{dRdr}.

Let us begin by eliminating the integral from Eq.\ \eqref{add} first as
\be{add3}
\text{arcsinh} \left\{D_A H_0^F(1+z)\sqrt{\Omega_k^F}\right\}=\sqrt{\Omega_k^F}\int_0^z\frac{dz}{\sqrt{\Omega_m^F(1+z)^3+\Omega_k^F(1+z)^2+\Omega_w^F(1+z)^w+\Omega_{\Lambda}^F}},
\ee
 and then differentiate both sides with respect $r$, yielding
\be{add4}
a_0 R'(r)+a_1 z'(r)=0,
\ee
where $a_0$ and $a_1$ are auxiliary and  $A(r)=R(r)/r$. We define a set of auxiliary functions $a_i$, $i=0,...9$, given below. 
\bea{Aa0}
a_0&=&\frac{H_0^F (z(r)+1)}{\sqrt{(H_0^F)^2 \Omega_k^F r^2
   A(r)^2 (z(r)+1)^2+1}}, \\ \label{Aa1}
a_1&=&\frac{H_0^F r A(r)}{\sqrt{(H_0^F)^2
   \Omega_k^F  r^2 A(r)^2 (z(r)+1)^2+1}}, \nonumber \\
&&-\frac{1}{\sqrt{\Omega_k^F
   (z(r)+1)^2-\Omega_k^F+\Omega_m^F (z(r)+1)^3-\Omega_m^F+\Omega_w^F
   (z(r)+1)^w-\Omega_w^F+1}}, \\
a_2&=&A(r), \\
a_3&=&\frac{\sqrt{\left(\frac{1}{A(r)}-1\right) \Omega (r)+1} \left[t(r)-t_b(r)\right] \left[2 \frac{dH_0(r)}{d\Omega(r)} (\Omega (r)-1) \Omega (r)+H_0(r) (\Omega (r)+2)\right]}{2 (\Omega (r)-1) \Omega (r)} \nonumber \\
&&-\frac{2 A(r)}{2 (\Omega (r)-1) \Omega (r)}, \\
a_4&=&-H_0(r)
   \sqrt{\left(\frac{1}{A(r)}-1\right) \Omega (r)+1}, \\
a_5&=&\sqrt{2 r^2 e(r)+1}, \\
a_6&=&H_0(r) (z(r)+1) \sqrt{\left(\frac{1}{A(r)}-1\right)
   \Omega (r)+1}, \\
a_7&=&\frac{z(r)+1}{4 A(r)^2 (\Omega(r)-1)}  \Bigg\{2 A(r)^2
    \bigg[2   \frac{dH_0(r)}{d\Omega (r)} (\Omega (r)-1)+H_0(r)\bigg]\sqrt{\left(\frac{1}{A(r)}-1\right) \Omega (r)+1} \nonumber \\
&& -H_0(r)   \left[t(r)-t_b(r)\right]\left[2 \frac{dH_0(r)}{d\Omega (r)} (\Omega (r)-1)
   \Omega (r)+H_0(r) (\Omega (r)+2)\right]\Bigg\} \\
a_8&=&\frac{H_0(r)^2 (z(r)+1) \Omega (r)}{2
   A(r)^2}, \\ \label{Aa9}
a_9&=&1-\frac{r H_0(r) \sqrt{\left(\frac{1}{A(r)}-1\right)
   \Omega (r)+1}}{\sqrt{2 r^2 e(r)+1}}.
\eea
We represent Eqs. (\ref{rs}), (\ref{ng}) and (\ref{dRdr}) using auxiliary functions $a_i$.   The common factors for all $a_i$ are that they all take a non-zero value at the origin and  they all depend only on $\Omega_M(r)$, $z(r)$, $t(r)$, $A(r)$ and $r$.
$H_0(r)$ is actually a function of $r$ through $\Omega_M(r)$ only, 
\be{AH0}
H_0(r)=H_{00} \left(\frac{1}{1-\Omega (r)}-\frac{\Omega (r) \text{arcsinh} \left(\sqrt{\frac{1-\Omega (r)}{\Omega (r)}}\right)}{(1-\Omega  (r))^{3/2}}\right).
\ee
With the help of the definitions \eqref{defs} and the gauge equation \eqref{gauge}, Eqs.\ \eqref{rs}, \eqref{ng} and \eqref{dRdr} now becomes
\bea{rs2}
z'(r)&=&\frac{a_6+(a_7 \Omega_M'(r)+a_8 t_b'(r))r}{a_5}, \\ \label{ng2}
t'(r)&=&\frac{a_2+(a_3 \Omega_M'(r)+a_4 t_b'(r))r}{a_5}, \\ \label{dRdr2}
R'(r)&=& a_9 [a_2 + (a_3  \Omega'(r) + a_4  t_b'(r)) r].
\eea
Taking the homogeneous bang time to be $t_b=0$, using the definition $R(r)=A(r) r$ and rearranging \eqref{add4}, \eqref{rs2},  \eqref{ng2} and \eqref{dRdr2} we obtain
\bea{Ars4}
z'(r)&=&\frac{a_0 (a_3 a_6 - a_2 a_7) a_9}{a_1 a_7 + a_0 a_3 a_5 a_9}, \\ \label{Ang4}
t'(r)&=&\frac{a_1 a_3 a_6 - a_1 a_2 a_7}{a_1 a_5 a_7 + a_0 a_3 a_5^2 a_9}, \\ \label{AdRdr4}
A'(r)&=&-a_2+\frac{a_1 (-a_3 a_6 + a_2 a_7) a_9}{a_1 a_7 + a_0 a_3 a_5 a_9}, \\ \label{AdOdr2}
\Omega_M'(r)&=&-\frac{a_1 a_6 + a_0 a_2 a_5 a_9}{a_1 a_7 r + a_0 a_3 a_5 a_9 r}.
\eea
This  set of differential equations   gives considerably more accurate solution than equations \eqref{rs}, \eqref{ng}, \eqref{add2} and \eqref{dRdr}.

\section{Solving the differential equations}\label{details}

\begin{figure}[t]
\centering
\includegraphics[scale=0.9]{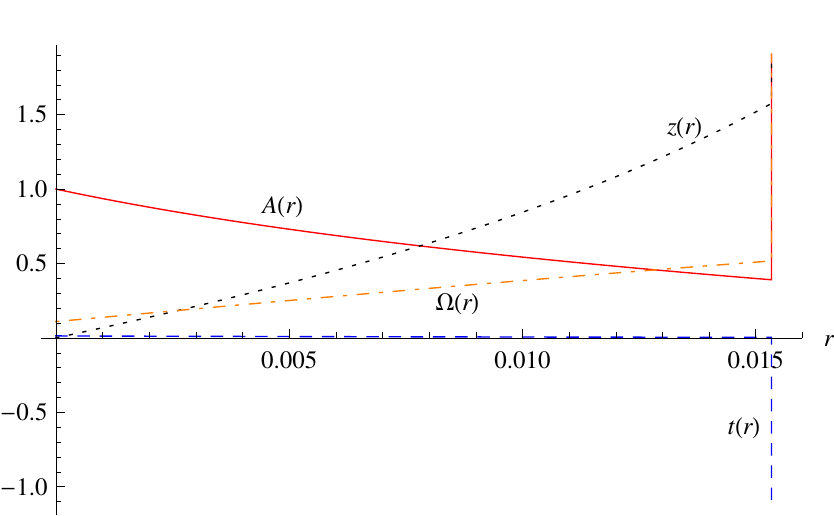} 
\includegraphics[scale=0.9]{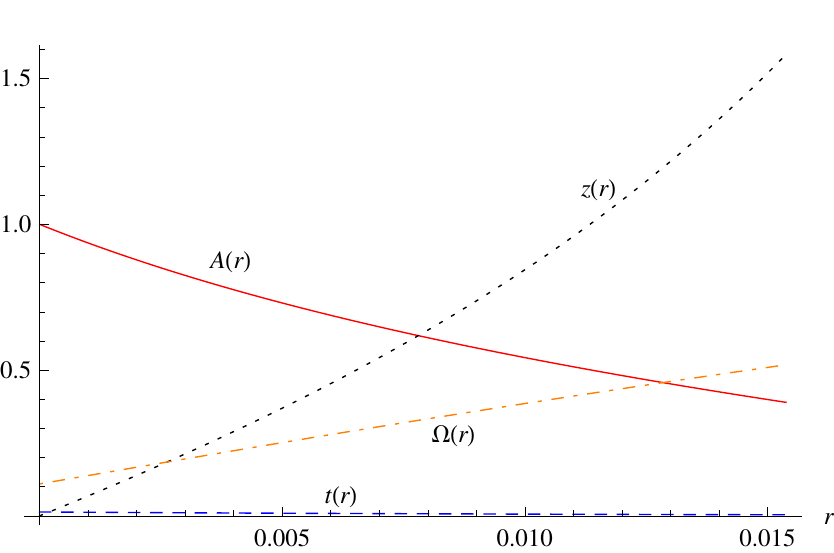} 
\caption{\footnotesize The  numerical evaluations for $\Omega_M$, $A$, $z$, and $t$ of the differential equations   \eqref{Ars4}-\eqref{AdOdr2} are plotted. In the left panel,   the evaluation begins at $r=0$ and ends at $r\approx 1.56545  \times 10^{-2}$, where the numerical procedure hits a singularity.  In the right panel, the evaluation begins at $r=r_{AH}$ and ends at $r\approx 6.84513 \times 10^{-9}$, where the numerical procedure hits a singularity. Colors available online. } 
\label{k1}
\end{figure}

 We follow the procedure to solve the system described in \cite{Krasinski2014}, but in a different gauge. We choose a gauge  which at the homogeneous limit approaches the conventional $\Lambda$CDM model,  making comparisons more straightforward. 
Because the   differential equations  \eqref{Ars4}-\eqref{AdOdr2} are not well-behaved close to the origin and the AH, we solve the system in pieces. In the vicinities of the origin and the AH, the system will be solved using linear approximations. 

In this appendix we  reconstruct the luminosity distance of the  \lcdm model given in  \cite{Krasinski2014}, where the  parameters of the homogeneous universe model are:
\be{fp's}
\Omega_m^F = 0.32, \quad  \Omega_k^F= 0, \quad \Omega_w^F = 0, \quad  H_0^F = 67.1.
\ee
The results are obtained similarly for different parameters.

\subsection{Solving the system starting from the origin towards the apparent horizon}

 Because the  set  \eqref{Ars4}-\eqref{AdOdr2} is singular at the origin, we use the linearized equations of the system from the origin to $r=10^{-6}$. From there on we use  Eqs.\ \eqref{Ars4}-\eqref{AdOdr2}. The initial conditions are solved in detail in Appendix \ref{AB}, which with our parametrisation yield the numerical values (with six digits accuracy)
\bea{oc's}
\begin{array}{ll}
&\Omega_M(0)=0.109621, \quad t(0)=0.0132964, \quad A(0)=1, \quad z(0)=0, \quad \\
& \Omega_M'(0)=28.5060, \quad t'(0)=-1, \quad A'(0)=-67.1000, \quad z'(0)=H_0^F.
\end{array}
\eea
Using these initial conditions, the solution of the differential equations  (\ref{Ars4})-(\ref{AdOdr2}) and its linearization are plotted in Figure \ref{k1}. The calculation stops in the vicinity of the AH, at $r\approx 0.016$, where the numerical methods report a singularity.

\subsection{Solving the system starting from the apparent horizon towards the origin}

 We use the linearized equations to calculate the quantities from $r_{AH}$ to $r_{AH}-10^{-6}$, where $r_{AH}$ is the comoving coordinate at the AH, and use full Eqs.\ \eqref{Ars4}-\eqref{AdOdr2} for smaller $r$. The relevant initial conditions  are   presented in  Appendix \ref{AAH} and their numerical values with six digits accuracy are
\bea{ahc's}
\begin{array}{ll}
&\Omega_{M,AH}=0.518126, \quad t_{AH}=0.00362764, \quad A_{AH}=0.389397, \quad z_{AH}=1.58243, \qquad  \\
& \Omega_{M,AH}'=23.4305, \quad t_{AH}'=-0.360968, \quad A_{AH}'=-25.3158, \quad z_{AH}'= 170.758,
\end{array}
\eea
where  the AH lies at $r=r_{AH}=0.0153816$ with six digit precision.   The solution is depicted in Figure \ref{k1}, where  functions $\Omega_M(r)$, $t(r)$, $A(r)$, and $z(r)$ are plotted. The numerical method hits  a singularity at the origin.

The solutions from the origin to the AH and from the AH to the  origin are glued together  at $r=0.0011$, where the absolute difference of each quantity between both solutions is less than $2 \times 10^{-6}$ and the relative differences are all less than $0.07 \%$.

\subsection{Solving the system beyond the apparent horizon}

\begin{figure}[t]
\centering
\includegraphics[scale=0.9]{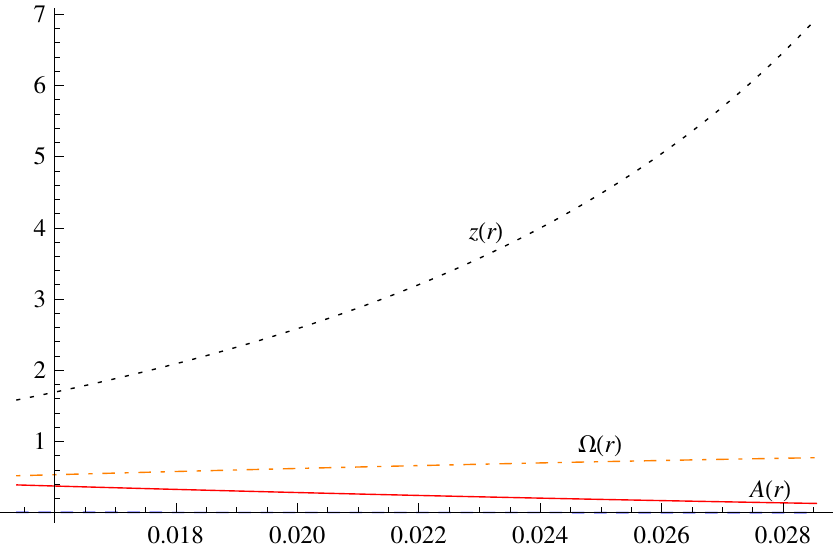} 
\caption{\footnotesize The numerical solutions for $\Omega_M$, $A$, $z$, and $t$ (dashed blue, overlapping with the horizontal axis)  of  DE's   \eqref{Ars4}-\eqref{AdOdr2} are plotted. The evaluation begins at $r=r_{AH}$ and ends at $r=r_{sc}$, where the system hits shell crossing singularity.  Colors available online.} 
\label{k7}
\end{figure}

 To go beyond the AH, we solve Eqs.\   \eqref{Ars4}-\eqref{AdOdr2} starting from $r=r_{AH}+10^{-6}$ towards infinity, using (\ref{ahc's}) as the initial conditions. The system hits the SC singularity at $r=r_{sc}\approx0.0285487$, corresponding to a redshift of $z=6.93850$,  which is the same value reported in \cite{Krasinski2014} within a precision of three digits. In Figure \ref{k7}, the solutions is drawn in the range $(r_{AH},r_{sc})$ and its linear approximation is drawn at the vicinity of the AH.

\section{The origin conditions} \label{AB}

To obtain a physically acceptable solution of the set  \eqref{Ars4}-\eqref{AdOdr2}, it has to satisfy  the relation \eqref{ah}. It is straightforward to see that the system is regular at the  origin, if $\lim_{r\rightarrow 0}R(r,t)=R_0(r)=r$, $\lim_{r\rightarrow 0}M(r)\propto r^3$, We can utilise the analysis of \cite{Krasinski2014}, where it was found that at the origin 
\be{B-2}
X^3+k X-2 M_0 H_0^F=0,
\ee
where $X=A_0(0) H_0^F$, $k=-2e(0)$ and $M_0=M(r)/r^3|_{r=0}$. Using $X$, $k$, and $M_0$, from Eqs.\ \eqref{R} and \eqref{eta} we find the present  age of the universe,
\be{B0}
t(0)=\frac{M_0}{(-k)^{3/2}}\left\{\sqrt{\left(\frac{-k}{M_0}\frac{X}{H_0^F}+1\right)^2-1}-\text{arccosh}\left(\frac{-k}{M_0}\frac{X}{H_0^F}+1\right)\right\},
\ee
from   Eqs.\ \eqref{B-2} and \eqref{defs} we get  $X^3[H_0(0)^2-(H_0^F)^2]=0$, and  Eqs.\ \eqref{defs} reveal
\be{B-3}
\Omega_M(0)=\frac{2 H_0^F}{X^3}.
\ee
It is straightforward to see that the values
\be{Xk}
H_0^F=67.1, \quad  M_0=100, \quad X= 49.6539 \quad \text{and} \quad k=-2195.24.
\ee
satisfy Eq.\ \eqref{B-2} and the system solved from the initial conditions obtained from these values agrees with the condition (\ref{ah}) at high precision.

Because we have defined $R(r,t)=A(r,t)r$, the  gauge \eqref{gauge} dictates  $A_0(0)=1$. Using the L'Hopital rule, one finds
\bea{B1}
\lim_{r\rightarrow0}\frac{R(r)}{r}&=&\lim_{r\rightarrow0} \frac{ \int_0^z \left[\Omega_m^F(1+z)^3+\Omega_k^F(1+z)^2+\Omega_w^F(1+z)^w+\Omega_{\Lambda}^F\right]^{-1/2}dz}{H_0^F r} \nonumber \\
&=&\lim_{r\rightarrow0} \frac{dz/dr}{H_0^F },
\eea
thus
\be{B2}
\frac{dz}{dr}\Big{|}_{r=0}=H_0^F.
\ee
To find  the value of $dt/dr$ in the origin, we combine  Eqs.\ \eqref{ng} and \eqref{dRdr} and use the chain rule, when
\be{B3}
R_{,r}(r,t)=\left( 1-\frac{R_{,t}(r,t)}{\sqrt{1+2 e(r)r^2}} \right)^{-1}\frac{d [(r(z),t(z)]}{dz}\frac{dz}{dr}.
\ee
On the other hand, differentiating  both sides of \eqref{add} with respect to $z$, yields
\bea{B4}
\frac{d R[r(z),t(z)]}{dz}& = &\left[\frac{1}{H_0^F(1+z)}+\frac{-z}{H_0^F(1+z)^2}\right] \frac{1}{\sqrt{\Omega_M^F(1+z)^3+1-\Omega_M^F}} +\mathcal{O}(z^2) \nonumber \\
&=&\frac{1}{H_0^F}+\frac{(-3 \Omega_M^F-4) z}{2
   H_0^F}+\mathcal{O}(z^2)
\eea
and, because  $H^T(0,t_0)=H_0^F$, Eq.\ \eqref{H2} yields
\be{B5}
R_{,t}(r,t)=H_0^F r+\mathcal{O}(z^2).
\ee
Inserting Eqs.\ \eqref{B2}-\eqref{B5} into Eq.\ \eqref{ng} gives 
\be{B6}
\frac{dt}{dr}\Big{|}_{r=0}= -1.
\ee
The values of $d\Omega(r)/dr$ and $dA(r)/dr$ at the origin we obtained by Taylor expansion of Eqs.\ \eqref{Ars4}-\eqref{AdOdr2} and finding the limit $r\rightarrow 0$. The result is
\be{b7}
\frac{d \Omega_M(r)}{dr} =-\frac{(H_0^F)^2 \left(\Omega_M(0)-1\right) \Omega_M(0) \left(3 \Omega _m-\Omega_M(0)-2\right)}{\left(2 \frac{d H_0(r)}{d \Omega_M(r)} \left(\Omega_M(0)-1\right) \Omega_M(0)+H_0^F
   \left(\Omega_M(0)+2\right)\right) \left(H_0^F t_0 \left(\Omega_M(0)+2\right)-2\right)},
\ee
\bea{b8}
\frac{d A(r)}{dr}&=&-H_0^F\Bigg{\{}\frac{H_0^F \left(3 \Omega _m+\Omega _M(r)+2\right) \left( \frac{d H_0(r)}{d \Omega_M(r)} t_0 \left(\Omega _M(r)-1\right) \Omega _M(r)-1\right)}{ \left(2  \frac{d H_0(r)}{d \Omega_M(r)} \left(\Omega
   _M(r)-1\right) \Omega _M(r)+H_0^F \left(\Omega _M(r)+2\right)\right) \left(H_0^F t_0 \left(\Omega _M(r)+2\right)-2\right)} \nonumber \\
&& +\frac{-8  \frac{d H_0(r)}{d \Omega_M(r)}
   \left(\Omega _M(r)-1\right) \Omega _M(r)+(H_0^F)^2 t_0 \left(\Omega _M(r)+2\right) \left(3 \Omega _m+\Omega _M(r)+2\right)}{2 \left(2  \frac{d H_0(r)}{d \Omega_M(r)} \left(\Omega
   _M(r)-1\right) \Omega _M(r)+H_0^F \left(\Omega _M(r)+2\right)\right) \left(H_0^F t_0 \left(\Omega _M(r)+2\right)-2\right)}. \nonumber \\
\eea

Finally, we note that the Eqs.\  \eqref{rs}, \eqref{ng}, \eqref{B2} and \eqref{B6} yield $H^R(0,t_0)=H_0^F$, and, as    $H^T(0,t_0)=H_0^F$, the mean Hubble value of the LTB model in the origin is  exactly same  as in the \lcdm model.

\section{Apparent horizon conditions} \label{AAH}

In this appendix we  mimic the luminosity distance of the  \lcdm model given in  \cite{Krasinski2014}  (the relevant  parameters are given in \eqref{fp's}). Similarly the results are obtained for different parameters.

 The redshift and the angular diameter distance at the AH are found where the derivative of the angular diameter distance \eqref{add} with respect to $z$ vanishes. The corresponding values are $z=z_{AH}\approx 1.58243$ and  $R[t(r_{AH}),r_{AH}]\approx 0.00598955$.
Numerical methods fail to evaluate Eqs.\  \eqref{Ars4}-\eqref{AdOdr2}  properly at the AH; this can be traced to Eq.\  \eqref{dRdr2}. By expanding $a_9$ in the vicinity of the AH, taking into account the AH conditions $dR[r,t(r)]/dr=0$ and \eqref{ah},  it can be written as $a_9=[a_{12}+a_{13}\Omega'(r)](r-r_{AH})+\mathcal{O}\left[(r-r_{AH})^2\right]$, where $a_{12}$ and $a_{13}$ are given in \eqref{a12} and \eqref{a13}.
On the other hand, we can solve $R'(r)$ from Eq.\ \eqref{add4} and expand it  around the  AH, yielding $R'(r)=a_{10} a_{11} z'(r)^2 (r-r_{AH})+\mathcal{O}\left[(r-r_{AH})^2\right]$, where $a_{10}$ and $a_{11}$ are given in \eqref{a10} and \eqref{a11}. Equating the expressions for $R'(r)$ we obtain
\be{Aah1}
a_{10} a_{11} z'(r)^2 =[a_{12}+a_{13}\Omega'(r)] [a_2 + (a_3  \Omega'(r) + a_4  t_b'(r)) r].
\ee 
The values for $A'(r)$, $\Omega'(r)$, $z'(r)$ and $t'(r)$ at the AH can now be solved from the equations \eqref{ah}, \eqref{rs2}, \eqref{ng2} and \eqref{Aah1} for the  linear expansion around the AH. Eq.\ \eqref{Aah1} shows that there are two solutions; we choose the one where $t'(r)<0$.

The functions $a_{10},...,a_{13}$ are given by
\bea{a10}
a_{10}&=&-\frac{\sqrt{ \Omega_k^F r^2 A(r)^2
   (z(r)+1)^2+1}}{ (z(r)+1)}, \\  \label{a11}
a_{11}&=&\frac{2 \Omega_k^F (z(r)+1)^2+3
   \Omega_m^F (z(r)+1)^3+\Omega_w^F w (z(r)+1)^w}{2 (z(r)+1)
   \left(\Omega_k^F (z(r)+1)^2-\Omega_k^F+\Omega_m^F
   (z(r)+1)^3-\Omega_m^F+\Omega_w^F
   (z(r)+1)^w-\Omega_w^F+1\right)^{3/2}} \nonumber \\ 
&&-\frac{(H_0^F)^3 \Omega_k^F
   r^3 A(r)^3 (z(r)+1)}{\left((H_0^F)^2 \Omega_k^F r^2 A(r)^2
   (z(r)+1)^2+1\right)^{3/2}}, \\ \label{a12}
a_{12}&=&\frac{3 r H_0(r)^2 \Omega (r)}{2
   A(r) \left[r^2 H_0(r)^2 (\Omega (r)-1)-1\right]}, \\ \label{a13}
a_{13}&=&\frac{r
   H_0(r) (2 r \frac{dH_0}{d\Omega } \Omega (r)+r
   H_0(r))}{2 A(r) \left[r^2 H_0(r)^2 (\Omega
   (r)-1)-1\right]}.
\eea

\section{The units} \label{AE}

In this article we use units where the speed of light $c=1$ and $c H_0^F=67.1$ km/(s $\times$ Mpc). This yields, within three digit precision,
\bea{units}
9.78 \times 10^{2} \,\text{Gy}&=&1, \\
3.00 \times 10^{2} \, \text{Gpc}&=& 1.
\eea
Thus,   the age of the LTB model is 13.0 Gy (when the Friedmannian parameters  \eqref{fp's} are assumed) and the cases where homogeneity is imposed at the drag epoch, void sizes are $\sim 10$ Gpc.


\begin{thebibliography}{X}



\bibitem{Riess1998}A. G. Riess et al., \emph{Observational Evidence from Supernovae for an Accelerating Universe and a Cosmological Constant}, Astron. J. 116, 1009 (1998), arXiv:astro-ph/9805201.

\bibitem{Perlmutter1999}S. Perlmutter et al., \emph{Measurements of $\Omega$ and $\Lambda$ from 42 High-Redshift Supernovae}, Astrophys. J. 517, 565 (1999),  	arXiv:astro-ph/9812133.




\bibitem{Celerier} Apparently, these models were originally intrduced in M.-N. C\'{e}l\'{e}rier, \emph{Do we really see a cosmological constant in the supernovae data ?}, Astron.Astrophys.353:63-71,2000,	arXiv:astro-ph/9907206,  for earlier void models, see J. W. Moffat at and D. C. Tatarski, \emph{Cosmological observations in a local void}, Astrophys. J. 453 (1995) 17 [arXiv:astro-ph/9407036].

\bibitem{VMC2012} W. Valkenburg, V. Marra ad  C. Clarkson, \emph{Testing the Copernican principle by constraining spatial homogeneity}, Mon.Not.Roy.Astron.Soc. 438 (2014) L6-L10, arXiv:1209.4078 [astro-ph.CO].

\bibitem{RedlichEtAl2014} M. Redlich, K. Bolejko, S. Meyer, G. F. Lewis, M. Bartelmann, \emph{Probing spatial homogeneity with LTB models: a detailed discussion}, A\&A 570, A63 (2014),  	arXiv:1408.1872 [astro-ph.CO].



\bibitem{Lemaitre} G. Lema\^{i}tre, Annales Soc. Sci. Brux. Ser. I Sci. Math. Astron. 
Phys. A 53 (1933) 51. For an English translation, see:
G. Lema\^{i}tre, \emph{The Expanding Universe}, Gen. Rel. Grav. 29 (1997) 641.

\bibitem{Tolman} R. C. Tolman, \emph{Effect Of Inhomogeneity On 
Cosmological Models}, Proc. Nat. Acad. Sci. 20 (1934) 169.

\bibitem{Bondi} H. Bondi, \emph{Spherically Symmetrical Models In General 
Relativity}, Mon. Not. Roy. Astron. Soc. 107 (1947) 410.

\bibitem{ClarksonCliftonFebruary2009} C. Clarkson, T. Clifton and S. February, \emph{Perturbation Theory in Lemaitre-Tolman-Bondi Cosmology}, JCAP 06 (2009) 025,  	arXiv:0903.5040 [astro-ph.CO].

\bibitem{AlnesAmarzguioui2006} H. Alnes and M. Amarzguioui, \emph{CMB anisotropies seen by an off-center observer in a spherically symmetric inhomogeneous universe},
Phys. Rev. D 74, 103520 (2006), arXiv:astro-ph/0607334v2.

\bibitem{BlomqvistMortsell2010}M. Blomqvist and E. M\"{o}rtsell, \emph{Supernovae as seen by off-center observers in a local void}, JCAP05(2010)006,  arXiv:0909.4723v2 [astro-ph.CO].

\bibitem{SundellVilja2014} P. Sundell and I. Vilja, \emph{Inhomogeneous cosmological models and fine-tuning of the initial state}, Mod. Phys. Lett. A  29, 1450053 (2014),  arXiv:1311.7290v2 [astro-ph.CO].


\bibitem{Garcia-BellidoHaugbolle2008a} J. Garcia-Bellido and T. Haugb\o lle, \emph{Confronting Lemaitre-Tolman-Bondi models with observational cosmology}, JCAP04(2008)003, arXiv:0802.1523v3 [astro-ph].


\bibitem{BNV2010} T. Biswas, A. Notari and W. Valkenburg, \emph{Testing the Void against Cosmological data: fitting CMB, BAO, SN and H0}, JCAP11(2010)030, arXiv:1007.3065v2 [astro-ph.CO].
 
\bibitem{ZG-BR-L2012} M. Zumalacarregui, J. Garcia-Bellido and P. Ruiz-Lapuente, \emph{Tension in the Void: Cosmic Rulers Strain Inhomogeneous Cosmologies}, JCAP10(2012)009, arXiv:1201.2790v3 [astro-ph.CO].

\bibitem{MossZibinScott2011} A. Moss, J. P. Zibin and D. Scott, \emph{Precision cosmology defeats void models for acceleration}, 	Phys. Rev. D83: 103515 (2011), arXiv:1007.3725v2 [astro-ph.CO].

\bibitem{Garcia-BellidoHaugbolle2008b} J. Garcia-Bellido and T. Haugb\o lle, \emph{Looking the void in the eyes—the kinematic Sunyaev–Zeldovich effect in Lemaître–Tolman–Bondi models}, JCAP09(2008)016,  	arXiv:0807.1326 [astro-ph].


\bibitem{ZibinMoss2011} J. P. Zibin and  A. Moss, \emph{Linear kinetic Sunyaev-Zel'dovich effect and void models for acceleration}, Class.Quant.Grav.28:164005,2011,  	arXiv:1105.0909 [astro-ph.CO].

\bibitem{ZhangStebbins2011} P. Zhang, A. Stebbins \emph{Confirmation of the Copernican principle at Gpc radial scale and above from the kinetic Sunyaev Zel'dovich effect power spectrum}, Phys.Rev.Lett.107:041301,2011,  arXiv:1009.3967v3 [astro-ph.CO].

\bibitem{BCF2012} P. Bull, T. Clifton and P. G. Ferreira, \emph{The kSZ effect as a test of general radial inhomogeneity in LTB cosmology}, Phys. Rev. D 85, 024002 (2012),  arXiv:1108.2222v3 [astro-ph.CO].





\bibitem{IguchiNakamuraNakao2002} H. Iguchi, T. Nakamura and K. Nakao, \emph{Is dark energy the only solution to the apparent acceleration of the present universe?},Prog.Theor.Phys. 108 (2002) 809-818, arXiv:astro-ph/0112419v2.

\bibitem{Krasinski2014} A. Krasi\'{n}ski, \emph{Accelerating expansion or inhomogeneity? Part 2: Mimicking acceleration with the energy function in the Lemaître-Tolman model}, Phys. Rev. D 90, 023524 (2014), arXiv:1405.6066v2 [gr-qc].

\bibitem{BKHC} K. Bolejko, A. Krasi\'{n}ski, C. Hellaby, and M.-N. C\'{e}l\'{e}rier, 
\emph{Structures in the Universe by Exact Methods}, Cambridge university press 
(2010).


\bibitem{HellabyLake1985} C. Hellaby, K. Lake, \emph{Shell crossings and the Tolman model}, Astrophys. J. 290, 381 (1985) + erratum Astrophys. J. 300, 461 (1985).


\bibitem{dePutterVerdeJimenez2012} R. dePutter, L. Verde, R Jimenez \emph{Testing LTB Void Models Without the Cosmic Microwave Background or Large Scale Structure: New Constraints from Galaxy Ages}, arXiv:1208.4534 [astro-ph.co] (2012).



\bibitem{BAO1}  F. Beutler, C. Blake, M. Colless, D. H. Jones, L. Staveley-Smith, et al., \emph{The 6dF Galaxy Survey: Baryon Acoustic Oscillations and the Local Hubble Constant}, Mon. Not. Roy. Astron. Soc. 416 (2011) 3017-3032,  	arXiv:1106.3366 [astro-ph.CO].


\bibitem{BAO2} N. Padmanabhan,  X. Xu, D. J. Eisenstein, R. Scalzo et al. \emph{A 2\% Distance to z=0.35 by Reconstructing Baryon Acoustic Oscillations - I : Methods and Application to the Sloan Digital Sky Survey},  	arXiv:1202.0090  (2012).


\bibitem{BAO3} L. Anderson, E. Aubourg, S. Bailey, D. Bizyaev, M. Blanton, et al., \emph{The clustering of
galaxies in the SDSS-III Baryon Oscillation Spectroscopic Survey: Baryon AcousticOscillations in the Data Release 9 Spectroscopic Galaxy Sample}, Mon. Not. Roy. Astron. Soc. 427 (2013), no. 4 3435-3467,  	arXiv:1203.6594 [astro-ph.CO]. 

\bibitem{BAO4}  C. Blake, E. Kazin, F. Beutler, T. Davis, D. Parkinson, et al., \emph{The WiggleZ Dark Energy Survey: mapping the distance-redshift relation with baryon acoustic oscillations}, Mon. Not. Roy. Astron. Soc. 418 (2011) 1707-1724,  	arXiv:1108.2635 [astro-ph.CO].

\bibitem{BAO5}T. Delubac, J. E. Bautista, N. G. Busca, J. Rich, D. Kirkby, et al., \emph{Baryon Acoustic Oscillations in the Ly$\alpha$ forest of BOSS DR11 quasars},A\&A 574, A59 (2015), arXiv:1404.1801v2 [astro-ph.CO]. 

\bibitem{BAO6} A. Font-Ribera, D. Kirkby, N. Busca, J. Miralda-Escudé, N. P. Ross, et al., \emph{Quasar-Lyman $\alpha$ Forest Cross-Correlation from BOSS DR11 : Baryon Acoustic Oscillations}, arXiv:1311.1767v2 [astro-ph.CO].


\bibitem{WMAP9yr} G. Hinshaw, D. Larson, E. Komatsu, D. N. Spergel, C. L. Bennett et. al. \emph{Nine-year Wilkinson microwave anisotropy probe (WMAP) observations: cosmological parameter results}, arXiv:1212.5226v3 (2013).

\bibitem{Planck2015} Planck Collaboration: P. A. R. Ade, N. Aghanim, M. Arnaud, M. Ashdown et. al., emph{Planck 2015 results. XIII. Cosmological parameters},  arXiv:1502.01589v2 [astro-ph.CO].

\bibitem{SN}M. Betoule, R. Kessler, J. Guy, J. Mosher, D. Hardin, R. Biswas et al., \emph{Improved cosmological constraints from a joint analysis of the SDSS-II and SNLS supernova samples}, arXiv:1401.4064v2 (2014).


\bibitem{Liddle2004} A. R. Liddle, \emph{How many cosmological parameters?}, Mon.Not.Roy.Astron.Soc. 351 (2004) L49-L53, arXiv:astro-ph/0401198v3.

\bibitem{Davis2007} T. M. Davis, E.  M\"{o}rtsell, J. Sollerman, A. C. Becker, S. Blondin et.al. \emph{Scrutinizing Exotic Cosmological Models Using ESSENCE Supernova Data Combined with Other Cosmological Probes},  	Astrophys.J.666:716-725,2007,  arXiv:astro-ph/0701510v2.


\bibitem{Schwarts1978} G. Schwarz, \emph{Estimating the Dimension of a Model}, The Annals of Statistics, 6, 461 (1978).

\bibitem{Plank} Planck Collaboration, P. Ade, N. Aghanim, C. Armitage-Caplan, M. Arnaud, M. Ashdown et al., \emph{Planck 2013 results. XVI. Cosmological parameters}, Astron. Astrophys. (2014),  arXiv:1303.5076v3 [astro-ph.CO]




\bibitem{ClarksonRegis2011} Chris Clarkson, and  Marco Regis,   \emph{The Cosmic Microwave Background in an Inhomogeneous Universe - why void models of dark energy are only weakly constrained by the CMB}, JCAP02(2011)013, arXiv:1007.3443v3.




\end{thebibliography}
\end{document}